\documentclass[aps,10pt,reprint,superscriptaddress,showpacs]{revtex4-2}

\usepackage{physics}
\usepackage{amssymb}
\usepackage{graphicx}
\usepackage{amsmath}
\usepackage{hyperref}
\usepackage{amsfonts,dsfont}

\graphicspath{{figures/}{sm_figures/}}

\begin{document}

\title{Realizing topologically ordered states on a quantum processor}

\newcommand{\xGoogle}{\affiliation{Google Quantum AI, Mountain View, CA, USA}}

\newcommand{\TUM}{\affiliation{Department of Physics, Technical University of Munich, 85748 Garching, Germany}}
\newcommand{\MCQST}{\affiliation{Munich Center for Quantum Science and Technology (MCQST), Schellingstr. 4, 80799 M{\"u}nchen, Germany}}
\newcommand{\IAS}{\affiliation{Institute for Advanced Study, Technical University of Munich, 85748 Garching, Germany}}

\newcommand{\Nottingham}{\affiliation{School of Physics and Astronomy, University of Nottingham, Nottingham, NG7 2RD, UK}}
\newcommand{\CQNE}{\affiliation{Centre for the Mathematics and Theoretical Physics of Quantum Non-Equilibrium Systems, University of Nottingham, Nottingham, NG7 2RD, UK}}

\newcommand{\caltech}{\affiliation{Department of Physics and Institute for Quantum Information and Matter, California Institute of Technology, Pasadena, CA, USA}}
\newcommand{\WBI}{\affiliation{Walter Burke Institute for Theoretical Physics, California Institute of Technology, Pasadena, CA, USA}}

\newcommand{\xUCSB}{\affiliation{Department of Physics, University of California, Santa Barbara, CA, USA}}
\newcommand{\xUMass}{\affiliation{Department of Electrical and Computer Engineering, University of Massachusetts, Amherst, MA, USA}}
\newcommand{\xMIT}{\affiliation{Research Laboratory of Electronics, Massachusetts Institute of Technology, Cambridge, MA 02139, USA}}
\newcommand{\Sorbonne}{\affiliation{Laboratoire de Physique Theorique et Hautes Energies, Sorbonne Universit\'e, France}}

\author{K. J. Satzinger} \xGoogle 
\author{Y. Liu} \TUM \MCQST
\author{A. Smith} \TUM \Nottingham \CQNE
\author{C. Knapp} \caltech \WBI

\author{M. Newman}  \xGoogle 
\author{C. Jones}  \xGoogle 
\author{Z. Chen}  \xGoogle 
\author{C. Quintana} \xGoogle
\author{X. Mi}  \xGoogle 
\author{A. Dunsworth}  \xGoogle 
\author{C. Gidney}  \xGoogle 

\author{I. Aleiner}  \xGoogle 
\author{F. Arute}  \xGoogle 
\author{K. Arya}  \xGoogle 
\author{J. Atalaya}  \xGoogle 
\author{R. Babbush}  \xGoogle 
\author{J. C. Bardin}  \xGoogle \xUMass
\author{R. Barends}  \xGoogle 
\author{J. Basso}  \xGoogle 
\author{A. Bengtsson}  \xGoogle 
\author{A. Bilmes}  \xGoogle 
\author{M. Broughton}  \xGoogle 
\author{B. B. Buckley}  \xGoogle 
\author{D. A. Buell}  \xGoogle 
\author{B. Burkett}  \xGoogle 
\author{N. Bushnell}  \xGoogle 
\author{B. Chiaro}  \xGoogle 
\author{R. Collins}  \xGoogle 
\author{W. Courtney}  \xGoogle 
\author{S. Demura}  \xGoogle 
\author{A. R. Derk}  \xGoogle 
\author{D. Eppens}  \xGoogle 
\author{C. Erickson}  \xGoogle 
\author{E. Farhi}  \xGoogle 
\author{L. Foaro}  \Sorbonne
\author{A. G. Fowler}  \xGoogle 
\author{B. Foxen}  \xGoogle 
\author{M. Giustina}  \xGoogle 
\author{A. Greene}  \xMIT \xGoogle 
\author{J. A. Gross}  \xGoogle 
\author{M. P. Harrigan}  \xGoogle 
\author{S. D. Harrington}  \xGoogle 
\author{J. Hilton}  \xGoogle 
\author{S. Hong}  \xGoogle 
\author{T. Huang}  \xGoogle 
\author{W. J. Huggins}  \xGoogle 
\author{L. B. Ioffe}  \xGoogle 
\author{S. V. Isakov}  \xGoogle 
\author{E. Jeffrey}  \xGoogle 
\author{Z. Jiang}  \xGoogle 
\author{D. Kafri}  \xGoogle 
\author{K. Kechedzhi}  \xGoogle 
\author{T. Khattar}  \xGoogle 
\author{S. Kim}  \xGoogle 
\author{P. V. Klimov}  \xGoogle 
\author{A.N. Korotkov}  \xGoogle 
\author{F. Kostritsa}  \xGoogle 
\author{D. Landhuis}  \xGoogle 
\author{P. Laptev}  \xGoogle 
\author{A. Locharla}  \xGoogle 
\author{E. Lucero}  \xGoogle 
\author{O. Martin}  \xGoogle 
\author{J. R. McClean}  \xGoogle 
\author{M. McEwen}  \xGoogle \xUCSB
\author{K. C. Miao}  \xGoogle 
\author{M. Mohseni}  \xGoogle 
\author{S. Montazeri}   \xGoogle 
\author{W. Mruczkiewicz}  \xGoogle 
\author{J. Mutus}
\author{O. Naaman}  \xGoogle 
\author{M. Neeley}  \xGoogle 
\author{C. Neill}  \xGoogle 
\author{M. Y. Niu}  \xGoogle 
\author{T. E. O'Brien}  \xGoogle 
\author{A. Opremcak}  \xGoogle 
\author{B. Pató}  \xGoogle 
\author{A. Petukhov}  \xGoogle 
\author{N. C. Rubin}  \xGoogle 
\author{D. Sank}  \xGoogle 
\author{V. Shvarts}     \xGoogle 
\author{D. Strain}  \xGoogle 
\author{M. Szalay}  \xGoogle 
\author{B. Villalonga}  \xGoogle 
\author{T. C.~White}  \xGoogle 
\author{Z. Yao}  \xGoogle 
\author{P. Yeh}  \xGoogle 
\author{J. Yoo}  \xGoogle 
\author{A. Zalcman}  \xGoogle 

\author{H. Neven}  \xGoogle 
\author{S. Boixo}  \xGoogle 
\author{A. Megrant}  \xGoogle 
\author{Y. Chen}  \xGoogle 
\author{J. Kelly}  \xGoogle 
\author{V. Smelyanskiy}  \xGoogle 
\author{A. Kitaev}  \xGoogle  \caltech \WBI
\author{M. Knap}  \TUM \MCQST \IAS
\author{F. Pollmann} \TUM \MCQST
\author{P. Roushan}  \xGoogle 

\begin{abstract}
The discovery of topological order has revolutionized the understanding of quantum matter in modern physics and provided the theoretical foundation for many quantum error correcting codes. Realizing topologically ordered states has proven to be extremely challenging in both condensed matter and synthetic quantum systems. Here, we prepare the ground state of the toric code Hamiltonian using an efficient quantum circuit on a superconducting quantum processor. We measure a topological entanglement entropy near the expected value of $\ln2$, and simulate anyon interferometry to extract the braiding statistics of the emergent excitations. Furthermore, we investigate key aspects of the surface code, including logical state injection and the decay of the non-local order parameter. Our results demonstrate the potential for quantum processors to provide key insights into topological quantum matter and quantum error correction.
\end{abstract}

\maketitle

%
%
Different phases of matter can commonly be distinguished in terms of spontaneous symmetry breaking and local order parameters. However, several exotic quantum phases have been discovered in recent decades that defy this simple classification, instead exhibiting \emph{topological order}~\cite{tsui:1982, WEN1990}. These phases are characterized by their long-range quantum entanglement and the emergence of quasiparticles with \emph{anyonic} exchange statistics. Moreover, they have energetically gapped ground states with degeneracies that depend on their boundary conditions. The non-local nature of these states makes them particularly attractive platforms for fault tolerant quantum computation, as quantum information encoded in locally indistinguishable ground states is robust to local perturbations~\cite{Kitaev2003,Freedman98}. This is the underlying principle of topological quantum error correcting codes, where the logical codespace corresponds to the degenerate ground state subspace of a lattice model~\cite{Bravyi1998, Freedman2001, Dennis02}.

An archetypical topological two-dimensional lattice model is the toric code, which exhibits so-called $\mathds{Z}_2$ topological order~\cite{Kitaev2003}. 
The realization of the toric code on a plane---the surface code---has emerged as one of the most promising stabilizer codes for quantum error correction due to its amenable physical requirements~\cite{Fowler2012}. Given both its inherent richness and quantum computing applications, experimentally realizing $\mathds{Z}_2$ topological order has sparked extensive interest, resulting in several experimental studies with comparatively small-scale synthetic quantum systems~\cite{Lu2009,Pachos2009,Feng2013,Park2016,Zhong2016,Dai2017,Luo2018,Song2018,Liu2019, Andersen2020,Erhard2021}.  Despite these efforts, the experimental realization of topologically ordered states remains a major challenge, requiring the generation of long-range entanglement. This can be achieved by identifying suitable quantum systems with topologically ordered ground states or by constructing a topologically ordered state in an engineered quantum system. Probing the non-local topological properties of such a state on an array of qubits requires high fidelity gates and a sufficiently large two-dimensional lattice.

%
%
\begin{figure*}[t!]
   \centering
   \includegraphics[width=490pt]{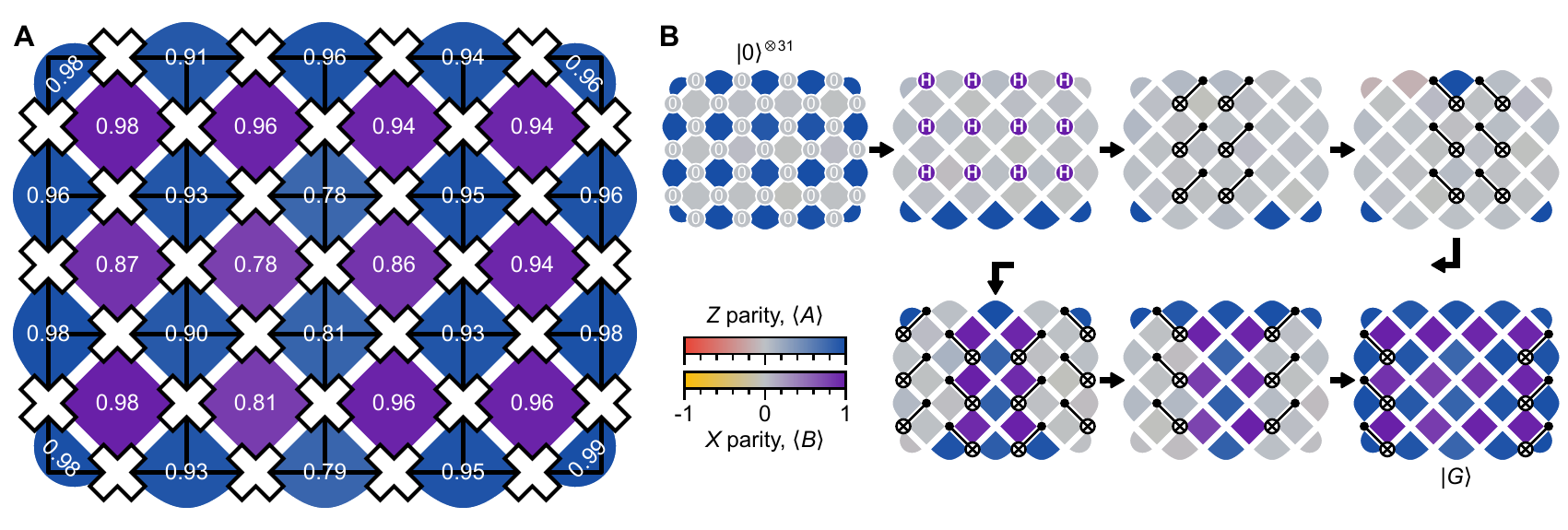}
   \caption{
   \textbf{Toric code ground state.} 
   (\textbf{A}) Experimentally-measured parity values for a 31-qubit lattice in the toric code ground state $\ket{G}$. Qubits ($\mathsf{x}$) are drawn on the standard toric code lattice, touching star ($A_s$, $+$, blue tile) and plaquette ($B_p$, $\square$, purple tile) operators.
   We compute each parity from a measured probability distribution (measuring each $A_s$ and $B_p$ separately, $10^4$ repetitions), which we correct for readout error using iterative Bayesian methods \cite{Nachman2020} (see \cite{SM} Sec.~B). Mean parity: $0.92\pm 0.06$ ($1\sigma$). 
   (\textbf{B}) Quantum circuit to prepare $\ket{G}$, with quantum gates superimposed on experimentally-measured parity values following each step. The circuit consists of Hadamard (H) and CNOT gates, which we compile into CZ gates.}
   \label{fig:1}
\end{figure*}

In this work, we develop an efficient quantum circuit to prepare the toric code ground state on a lattice of 31 superconducting qubits. We then experimentally establish the topological nature of the state by measuring the topological entanglement entropy. By simulating interferometry of toric code excitations, we fully determine their associated braiding statistics. Furthermore, we prepare logical states of the distance-5 surface code on 25 qubits and demonstrate error correction of logical measurements. While a meaningful implementation of active error correction on these states is beyond current experimental capabilities, we realize these states without stabilizer circuitry, providing a scheme to characterize and understand errors of logical qubits.

We realize the toric code ground state, depicted in Fig.~\ref{fig:1}A, by implementing a shallow quantum circuit on a Sycamore quantum processor \cite{Arute2019a}.
The toric code Hamiltonian
\begin{equation}\label{eq:TC}
    H = -\sum_{s} A_{s} - \sum_{p} B_{p}
\end{equation}
is defined in terms of qubits living on the edges of a square lattice.
The ``star" operators $A_s = \prod_{i\in s} Z_i$ are products of Pauli $Z$ operators touching each star ($+$, blue). The ``plaquette" operators $B_p = \prod_{j \in p} X_j$ are products of Pauli $X$ operators on each plaquette ($\square$, purple).
For the boundary conditions shown in Fig.~1A, there is a unique toric code ground state $\ket{G}$, with parity $+1$ for all star and plaquette operators: $A_s\ket{G} = B_p\ket{G} = +1\ket{G}$.

Our ground state preparation algorithm, depicted in Fig.~\ref{fig:1}B, is motivated by the observation that the ground state is an equal superposition of all possible ``plaquette configurations" and can be written as
\begin{equation}
    \ket{G} = \frac{1}{\sqrt{2^{12}}}\prod_{p} (\mathbb{I}+B_p)\ket{0}^{\otimes 31},
    \label{eq:GS_TC}
\end{equation}
where $\ket{0}^{\otimes 31}$ is the product of single-qubit states $\ket{0}$, and the product is over the 12 plaquettes. We begin in the trivial state $\ket{0}^{\otimes 31}$, where all $\langle A_s\rangle = 1$ and $\langle B_p\rangle = 0$. For each plaquette $B_p$, we perform a Hadamard on the upper qubit, preparing $(\ket{0}+\ket{1})/\sqrt{2}$, and then perform CNOT gates to the other qubits on the plaquette, effectively realizing $\mathbb{I}+B_p$. These operations are carefully ordered, starting in the middle and working outward, to avoid conflict between plaquettes while minimizing circuit depth. The 12 Hadamards create a superposition of $2^{12}$ bitstrings, and the CNOTs transform each of those bitstrings into a configuration where the $Z$ parity on each star is $+1$; the final superposition has $X$ parity $+1$ on each plaquette. This circuit exhibits optimal scaling, with depth \emph{linear} in system width \cite{Bravyi2006}, specifically $3+2\lfloor(N-1)/2\rfloor$ nearest-neighbor CNOT layers for a lattice $N$ plaquettes wide.

%
%
\begin{figure*}[t!]
   \centering
   \includegraphics[width=342pt]{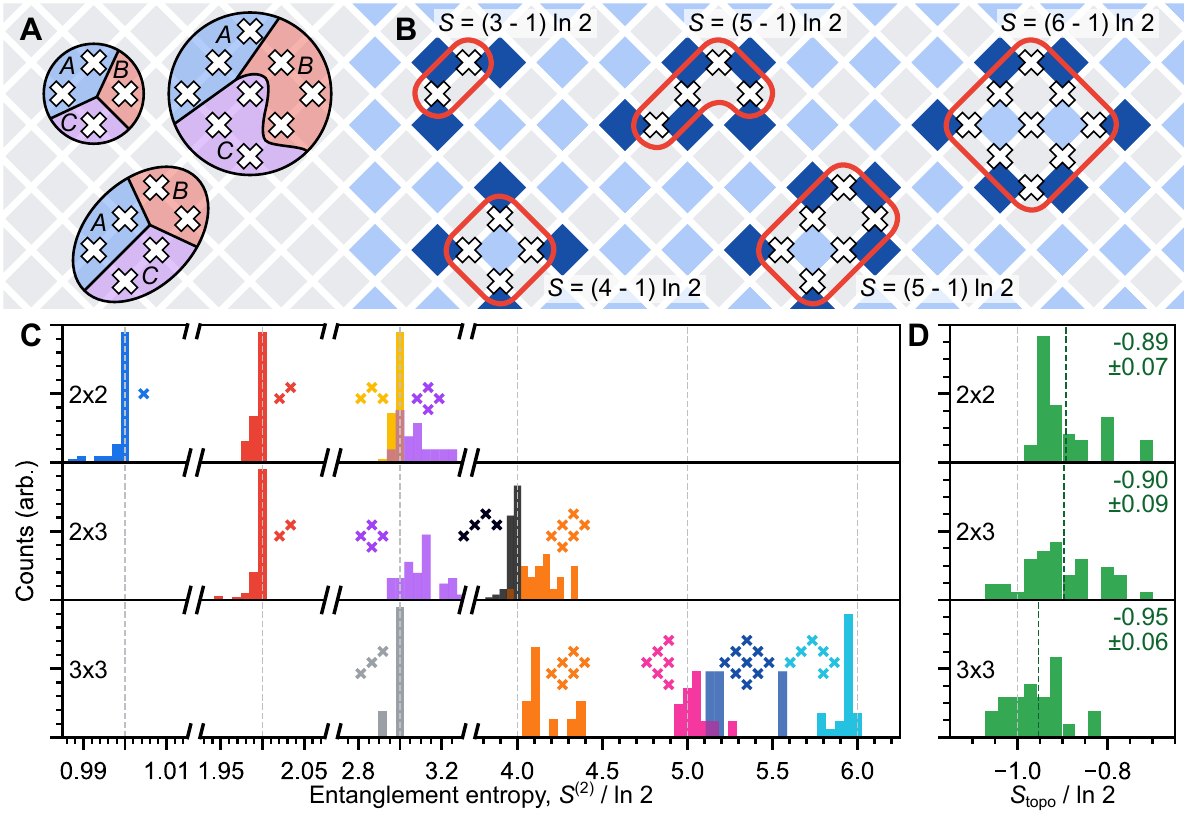}
   \caption{
   \textbf{Topological entanglement entropy.}
   (\textbf{A}) Schematic identifying the subsystems $A$, $B$, and $C$ used to measure topological entanglement entropy $S_\textrm{topo}$ on 4-, 6-, and 9-qubit systems within the toric code lattice.
   (\textbf{B}) Illustration identifying the expected entanglement entropy $S$ for groups of qubits in the toric code. We draw a red perimeter around each group and count the number $k$ of star operators (blue tiles) it crosses. $S = k\ln 2 + S_\text{topo} = (k-1)\ln 2$.  
   (\textbf{C}) Experimental second R\'enyi entropy $S^{(2)}$ distributions measured on the 31-qubit toric code ground state.
   There is a histogram for each subsystem shape.
   (\textbf{D}) Topological entanglement entropy $S_\text{topo}/\ln2$ (ideal value $-1$) computed from the entropies in \textbf{C}. We evaluate each dataset in all possible orientations of the subsystems in \textbf{A} ($2\times 2$: 4, $2\times 3$: 2, $3\times 3$: 8).
   Upper right: mean (dark green line) and distribution standard deviation.
   }
   \label{fig:2}
\end{figure*}

Topologically ordered states in 2D systems exhibit long-range quantum entanglement, characterized by the topological entanglement entropy, $S_\text{topo}$~\cite{Kitaev06,Levin2006}. Ground states of 2D gapped Hamiltonians typically satisfy the ``area law" scaling of the entanglement entropy: the leading-order contribution to the entanglement entropy $S_A$ of a subsystem $A$ results from local interactions that scale with the boundary length of the subsystem. Topological ground states have an additional universal constant contribution $S_\text{topo}<0$ arising from their intrinsic long-range entanglement. 
To extract $S_\text{topo}$, a linear combination of subsystem entropies can be constructed such that the local contributions cancel. For the subsystems depicted in Fig.~\ref{fig:2}A,
\begin{align}\label{eq:Stopo}
S_\text{topo} & = S_A + S_B + S_C - S_{AB} - S_{BC} - S_{AC} + S_{ABC},
\end{align}
where $AB$ indicates the union of $A$ and $B$. The structure of toric code eigenstates implies that $S_\text{topo}$ can be inferred from small subsystems. For the toric code ground state, $S_\text{topo} = -\ln 2$,  reflecting the total quantum dimension of $\mathds{Z}_2$ topological order~\cite{Kitaev2006b},  
while $S_\text{topo} =0$ in the absence of topologically order. 

The structure of the toric code Hamiltonian results in entanglement characterized by integer multiples of $\ln2$, scaling with the number of star operators $A_{s}$ intersecting the subsystem boundary~\cite{Hamma2005}, as illustrated in in Fig.~\ref{fig:2}B.  To compute $S_\text{topo}$, one can measure the second R\'enyi entropy  $S^{(2)} = -\ln[\Tr(\rho^2)]$, where $\rho$ is density matrix, for each subsystem in Eq.~\eqref{eq:Stopo}. 
Recently-introduced randomized methods enable efficient measurement of R\'enyi entanglement entropies, requiring a smaller number of measurements for large subsystems compared to full quantum state tomography~\cite{vanEnk2012, Elben2018, Brydges2019}. This enables accurate entropy measurement when tomography is intractable, such as the 9-qubit subsystem in Fig.~\ref{fig:2}A.
We apply random single-qubit unitaries to the subsystem of interest and sample the probability distribution of the bitstrings.
Analyzing statistical correlations across many random instances allows us to compute the second R\'enyi entropy. We use an iterative Bayesian scheme \cite{Nachman2020} to mitigate measurement errors and remove under-sampling bias (see \cite{SM} Sec.~C, where we also compare randomized measurement with tomography results).

Figure~\ref{fig:2}C shows distributions of the measured entanglement entropies for subsystems of $2\times 2$, $2\times 3$, and $3\times 3$ qubits within the toric code ground state. 
For a subsystem with $n$ qubits, the entanglement entropy ranges from $0$ for a product state up to $n\ln 2$. In the toric code, subsystems with no interior have the maximum value $n\ln 2$; 
in those cases, we measure a narrow distribution centered just below the ideal value. For subsystems with an interior, we measure a wider distribution centered slightly above the predicted value. This is consistent with unitary error and decoherence slightly mixing the system with its environment, which increases entanglement entropies that are not yet at their maximal value.

We compute $S_\text{topo}$ from the subsystem entropies using Eq.~\eqref{eq:Stopo} for 14 different $2\times 2$ arrays, 20 different $2\times 3$ arrays, and 3 different $3\times 3$ arrays. Each randomized measurement on the qubit array yields several $S_\text{topo}$ estimates from different orientations of the partitions $A, B, C$. Distributions of measured $S_\text{topo}$ are shown in Figure~\ref{fig:2}D, with mean values $S_\text{topo}/\ln2=-0.89$, $-0.90$, and $-0.95$ for the $2\times 2$, $2\times 3$, and $3\times 3$ qubit arrays, respectively. The distributions provide strong evidence for the non-trivial topological nature of the state, closely approaching the ideal value of $S_\text{topo}=-\ln2$, completely distinct from the trivial state value of zero.

%
%
\begin{figure*}[t!]
   \centering
   \includegraphics[width=342pt]{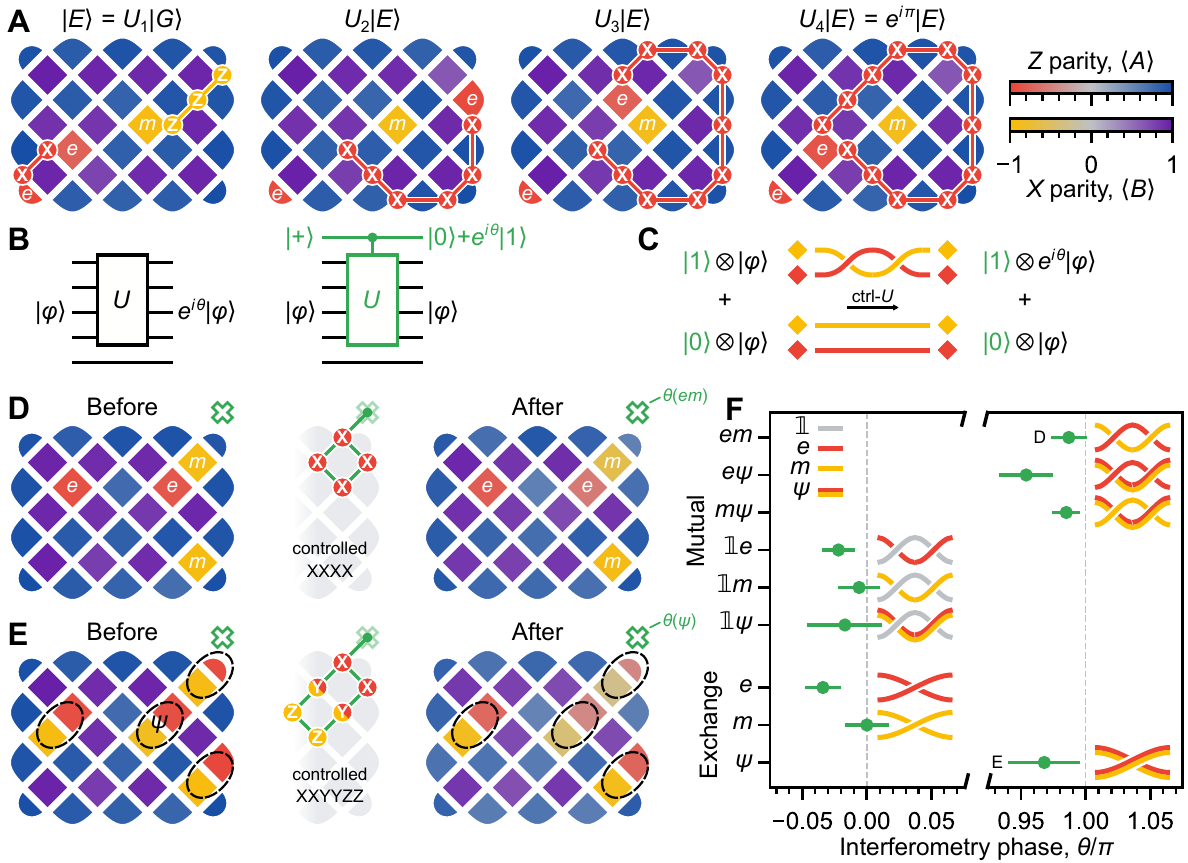}
   \caption{
   \textbf{Extracting braiding statistics using Ramsey interferometry.} 
   (\textbf{A}) Visualizing braiding with a toric code excited state $|E\rangle=U_1|G\rangle$ (excitations $e$ (red) and $m$ (yellow), experimentally-measured parities). We apply additional $X$ gates ($U_2$, $U_3$, $U_4$) to visualize braiding an $e$ around the $m$. 
   (\textbf{B}) Quantum circuits with unitary $U$ and an eigenstate $\ket{\varphi}$. Left: Direct application. Right: Extracting the phase $\theta$ using an auxiliary qubit (green).
   (\textbf{C}) Illustration of Ramsey interferometry for the case of braiding an $e$ and $m$ (state $\ket{\varphi}$) using an operator $U$. We visualize the superposition of two paths, with the braid operation $U$ controlled by an auxiliary qubit in $\ket{{+}}$.
   (\textbf{D}) Extracting the mutual statistics for $e$ and $m$. Left: initial excited eigenstate (similar to \textbf{A}). We implement controlled-\textit{XXXX} with an auxiliary control qubit (green) starting in $|+\rangle$. Right: parity measurements after controlled-\textit{XXXX}.
   (\textbf{E}) Extracting the fermion exchange statistics, analogous to \textbf{D}. We create two pairs of $\psi$ (neighboring $e$ and $m$) and implement controlled-\textit{XXYYZZ} to measure the exchange phase. 
   (\textbf{F}) Measured mutual and exchange phases, with braiding diagrams. Phases are from tomography on the auxiliary qubit, 18000 total repetitions per compiled instance. Standard error estimated with jackknife re-sampling over instances.
   }
   \label{fig:3}
\end{figure*}

One of the most exotic features of topological phases is that their quasiparticle excitations (anyons) satisfy mutual and exchange statistics more general than those of bosons and fermions. Excitations in toric code eigenstates are commonly denoted as ``electric" $e$ with $\langle A_s \rangle = -1$, and ``magnetic" $m$ with $\langle B_p \rangle = -1$, in connection to lattice gauge theory. The four distinct anyons of the toric code are $\mathds{1}$ (the absence of an $e$ or $m$), $e$, $m$, and $\psi$ (an emergent fermion resulting from the combination of $e$ and $m$). In the toric code, the mutual statistics are encoded in the phase accumulated when dragging one anyon around another anyon of different type, while the exchange statistics are phases arising from spatial interchange of two identical anyons. We simulate anyon braiding on our quantum processor by creating toric code excited states corresponding to all the distinct excitations and measuring their statistics with interferometry.

The toric code excited states can be created by applying a string of Pauli operators to the ground state: an $X$-string will result in the state with $e$ excitations at each end, while a $Z$-string prepares the state with $m$ excitations at each end, as shown in the first panel of Fig.~\ref{fig:3}A, where we visualize an example of $e-m$ mutual braiding with snapshots of experimentally-measured parity values, $\langle A_s\rangle$ and $\langle B_p\rangle$. 
We move an $e$ around $m$ with an $X$-string, eventually returning to its initial position. The initial and final states have the same parity values but differ by an overall phase, in this case $\pi$, which is not directly detectable.

To experimentally extract the braiding statistics, we employ a multi-qubit Ramsey interferometry scheme~\cite{Jiang2008}. This protocol provides a scalable way to measure the overlap between the initial and final states, allowing experimental access to the accumulated phase $\theta$. A key step in this protocol is the use of an auxiliary qubit and a controlled operation, effectively creating a superposition of the braided and non-braided states, as shown in Fig.~\ref{fig:3}B-C. This sequence imparts $\theta$ into a measurable rotation of the auxiliary qubit. We efficiently compile the multi-qubit controlled operations into CZ gates. Since the measured phases are sensitive to coherent and non-Markovian errors, we use randomized compiling to mitigate these errors \cite{Wallman2016}. See Sec.~D in \cite{SM} for details.

%
%
\begin{figure*}[t!]
   \centering
   \includegraphics[width=342pt]{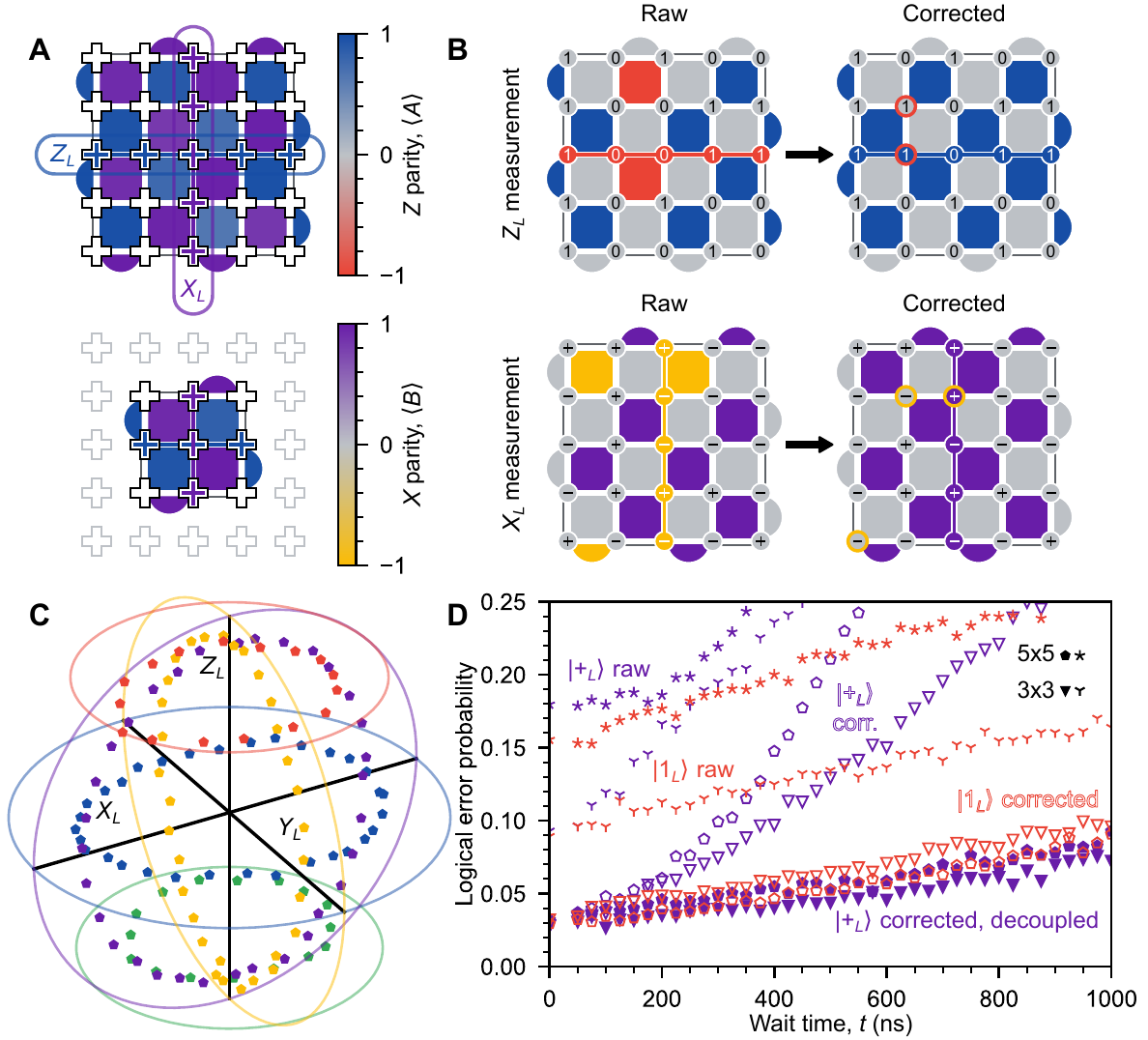}
   \caption{
   \textbf{Surface code logical qubit states.} 
   (\textbf{A}) Measured parity values for surface code logical qubit states $\ket{T_L}=\left(\ket{0_L}+e^{i\pi/4}\ket{1_L}\right)/\sqrt{2}$ on $5\times 5$ and $3\times 3$ qubit arrays. Logical operators $Z_L$ and $X_L$ span across each array.
   (\textbf{B}) Logical measurement with error correction. We measure a 25-qubit bitstring in $X$ or $Z$ basis and evaluate the local parities of the same basis. Negative parities indicate an error. We flip the circled qubits to restore positive parities.
   (\textbf{C}) Experimental logical qubit tomography immediately after state injection for 128 states (sweeping the initial state of the center qubit $\ket{\alpha}+\ket{\beta}$), plotted in the Bloch sphere ($5\times 5$). The ideal states lie on five planes: $x=0$ (yellow),  $y=0$ (purple), $z=1/\sqrt{2}$ (red), $z=0$ (blue), $z=-1/\sqrt{2}$ (green). 
   Mean Bloch vector length: $0.6\pm0.1$ ($1\sigma$).
   (\textbf{D}) We prepare logical states, wait for a time $t$, and then perform a logical measurement.
   For $|1_L\rangle$, we compare raw (hollow markers) to corrected (filled markers) logical measurements. 
   For $|+_L\rangle$, we compare free evolution (filled markers) to dynamical decoupling (star markers), both using corrected logical measurement (see main text).
   Each logical measurement uses $10^4$ repetitions.
   }
   \label{fig:4}
\end{figure*}

Fig.~\ref{fig:3}D-E illustrate two examples of braiding interferometry. In Fig.~\ref{fig:3}D, we extract the $e-m$ mutual statistics, where the braiding path is a Pauli string \textit{XXXX}, moving $e$ around the plaquette that contains an $m$. Fig.~\ref{fig:3}E shows a similar example for the exchange statistics of two identical $\psi$ excitations using a path of intertwining Pauli strings of \textit{XXXX} and \textit{ZZZZ}, simplifying to \textit{XXYYZZ} (see Sec.~D in \cite{SM} for details). The parity measurements show consistent values before and after the controlled-braiding operation, slightly fading due to decoherence and gate error. We measure the phases for the other mutual and exchange combinations, presenting the results in Fig.~\ref{fig:3}F. The phases are plotted alongside their corresponding braid diagrams, with the expected values $0$ and $\pi$ indicated by dashed gray lines.

Our measurements illuminate the non-trivial mutual and exchange statistics of the toric code. 
Braiding $e$ around $m$ results in a $\pi$ phase, which does not occur for local bosons or fermions. Moreover, while $e$ and $m$ both satisfy bosonic exchange statistics, their combination $\psi$ exhibits fermionic exchange statistics.  The mutual and exchange statistics of the anyons, conventionally summarized in the modular $\mathcal{S}$ and $\mathcal{T}$ matrices, fully characterize the $\mathds{Z}_2$ topological order~\cite{Kitaev2006b}.

Distinct topologically ordered ground states are locally indistinguishable, making them attractive logical qubits due to this immunity to local perturbations. The lattice of Fig.~1A has only one ground state, but in Fig.~4A we use different boundary conditions where the toric code admits a ground state degeneracy, as proposed for the surface code~\cite{Bravyi1998, Freedman2001, Horsman2012}. We introduce \emph{logical operators} $Z_L$ and $X_L$ which span across the lattice and commute with the Hamiltonian but anti-commute with each other.

We generalize the state preparation circuit of Fig.~1B to create the logical states $\ket{0_L}$ and $\ket{+_L}$, where $Z_L\ket{0_L}=+1\ket{0_L}$ and $X_L\ket{{+}_L}=+1\ket{{+}_L}$, on both $5\times 5$ (distance-5) and $3\times 3$ (distance-3) arrays. The $\ket{0_L}$ and $\ket{{+}_L}$ preparations are closely related, connected by a logical Hadamard. We then use the logical operators, which are simply products of single-qubit gates, to realize $\ket{1_L}=X_L\ket{0_L}$ and $\ket{{-}_L}=Z_L\ket{{+}_L}$. See Sec.~A in \cite{SM} for details on state preparation and logical operations.

The logical states are resilient to local errors, which we demonstrate with logical measurement with error correction, shown in Fig.~4B. Following surface code proposals, we perform a logical measurement by projectively measuring all the qubits in $Z$ or $X$ basis (for $Z_L$ or $X_L$, respectively). Naively evaluating the parity of the logical operator is vulnerable to errors on any qubit along the operator, but errors can be detected by also evaluating the local parities ($A_s$ and $B_p$) from the individual qubit measurements. By construction, we expect the local parities to be $+1$, so any $-1$ values indicate nearby errors. We find a minimal set of qubits to flip in order to recover $+1$ parities before evaluating the logical operator. This correction decreases the logical error by about a factor of 5 for distance-5 and a factor of 3 for distance-3. Averaging over $Z_L$ and $X_L$ eigenstates, the logical preparation and measurement error is 0.030 for both distance-5 and distance-3, lower than the average physical qubit preparation and measurement error, 0.034.
This is a simplified form of error correction compared to the repetitive stabilizer measurements of surface code proposals, where parity changes are matched together over space and time.

The logical subspace also admits arbitrary superposition states $\alpha\ket{0_L}+\beta\ket{1_L}$, which we realize with \emph{state injection}, encoding a single physical qubit state into the logical qubit. For $5\times 5$ state injection, we prepare the central qubit in $\alpha\ket{0}+\beta\ket{1}$ and then create a GHZ-like state $(\alpha\mathbb{I}+\beta X_L)\ket{0}^{\otimes 25}$ using three CZ layers. The toric code preparation circuit maps $\ket{0}^{\otimes 25}\to \ket{0_L}$ and $X_L\ket{0}^{\otimes 25}\to \ket{1_L}$, giving $\alpha\ket{0_L}+\beta\ket{1_L}$. For example, the states depicted in Fig.~4A are logical $T$ states $\ket{T_L}=\left(\ket{0_L}+e^{i\pi/4}\ket{1_L}\right)/\sqrt{2}$, of interest for non-Clifford operations. We characterize injected states using logical tomography. Measuring $Z_L$ and $X_L$ is straightforward and robust, as discussed above. We measure $Y_L$ by performing another logical gate, $X_L^{1/2}=(\mathbb{I}-iX_L)/\sqrt{2}$, decomposed into five CZ layers, and then measuring $Z_L$. We plot the resultant Bloch vectors for 128 injected states across the Bloch sphere in Fig.~4C. By measuring these non-local order parameters, we illuminate the logical degree of freedom that was invisible to the local parity measurements of Fig.~4A.

Finally, we investigate decoherence of $Z_L$ and $X_L$ eigenstates by plotting logical error versus wait time $t$ in Fig.~4D.
We reiterate the importance of measurement error correction by comparing raw and corrected data for $\ket{1_L}$. Note that while distance-5 has significantly worse raw error, after correction it is modestly better than distance-3. However, observe that $\ket{{+}_L}$ decays much more quickly than $\ket{1_L}$, due to its sensitivity to $Z$ errors (dephasing). We dynamically decouple the qubits from low-frequency noise with a simple sequence executing an $X$ gate on each qubit at $t/4$ and $3t/4$, which brings $\ket{{+}_L}$ error slightly below $\ket{1_L}$ error, with distance-3 remaining slightly lower-error. $\ket{1_L}$ and $\ket{0_L}$ are not appreciably affected by this dynamical decoupling (see Sec.~E in \cite{SM}). 
Overall, the logical error increases linearly at 0.06 per microsecond. For active error correction with the surface code, we expect a few percent logical error per cycle at threshold~\cite{Fowler2012}. Typical cycle durations are hundreds of nanoseconds~\cite{Chen2021}, where the logical state suffers the decoherence studied here as well as gate errors, suggesting continued efforts to decrease the cycle duration and improve coherence.

%
%
Our shallow quantum circuits for realizing toric code eigenstates can be extended to other topologically ordered states, including string-nets with non-Abelian anyons~\cite{Levin2005}.  Moreover, the tools we have developed can be readily applied to a wide class of topologically-ordered states generated on quantum processors. By encoding quantum information in the degenerate ground state manifold of the toric code, we provide a method for studying coherence properties of logical qubit states. This method could be used to identify and mitigate noise correlations in the system, with critical implications for future error correction experiments.

%
%
\vspace{3mm}
\textbf{Acknowledgements.}  We thank B. Bauer, A. Elben, B. Vermersch, and G. Vidal for useful discussions.
FP, YL, AS, and MK acknowledge support from the Technical University of Munich - Institute for Advanced Study, funded by the German Excellence Initiative and the European Union FP7 under grant agreement 291763, the Max Planck Gesellschaft (MPG) through the International Max Planck Research School for Quantum Science and Technology (IMPRS-QST), the Deutsche Forschungsgemeinschaft (DFG, German Research Foundation) under Germany’s Excellence Strategy--EXC--2111--390814868, TRR80 and DFG grant No. KN1254/2-1, and from the European Research Council (ERC) under the European Union’s Horizon 2020 research and innovation programme (grant agreements No. 771537 and No. 851161).
AS was supported by a Research Fellowship from the Royal Commission for the Exhibition of 1851.
CK was supported by the Walter Burke Institute for Theoretical Physics at Caltech, and by the IQIM, an NSF Frontier center funded by the Gordon and Betty Moore Foundation, the Packard Foundation, and the Simons Foundation.

\textbf{Correspondence}. All correspondence should be addressed to F.~Pollmann (frank.pollmann@tum.de) and P.~Roushan (pedramr@google.com). 

\newpage

\bibliography{references.bib}

%
%
\newcommand{\beginsupplement}{
  \setcounter{table}{0}  
  \renewcommand{\thetable}{S\arabic{table}} 
  \setcounter{figure}{0} 
  \renewcommand{\thefigure}{S\arabic{figure}}
  \setcounter{section}{0}
  \setcounter{equation}{0}
  \renewcommand{\theequation}{S\arabic{equation}}
}

\beginsupplement

\newpage
\onecolumngrid
\begin{center}
	\textbf{\large Supplementary materials for\\
	``Realizing topologically ordered states on a quantum processor"}
\end{center}
\bigskip
\twocolumngrid

\section{Linear quantum circuit for the toric code}

We provide a general circuit design principle for the toric code on a square lattice. The construction can be implemented on Sycamore to (i) realize toric code with $3+2\lfloor(N-1)/2\rfloor$ nearest-neighbor (NN) CNOT layers for a lattice with $N\times M$ plaquettes, where $N\leq M$, and (ii) encode arbitrary distance-$d$ logical qubit with $d+1$ layers of nearest-neighbour CNOT (however, in some instances this reduces to $(d+3)/2$ layers). Such construction is generalizable to a wide range of Abelian and non-Abelian quantum codes~\cite{Liu2021}. The linear scaling of the circuit is essentially optimal for topologically ordered states \cite{Bravyi2006}. 

\begin{figure}
    \centering
    \includegraphics{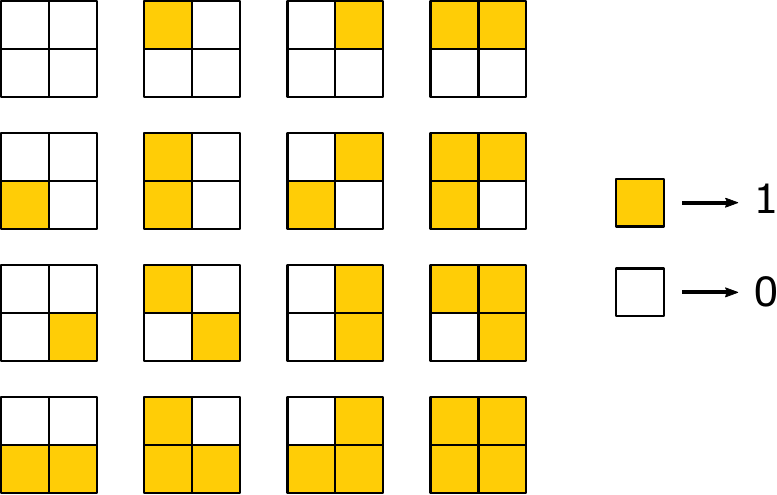}
    \caption{The binary correspondence of the configurations (toric code on $2\times 2$ plaquette system). The yellow plaquettes are acted by $B_p$, the white plaquettes are acted by $\mathbb{I}$. This can be viewed as an equal-weight superposition of binary strings where the binary digit corresponds to the two operators.}
    \label{sm:fig:binary}
\end{figure}

As mentioned in the main text, a toric code ground state takes the form of a product of commuting projectors 
\begin{equation}
\ket{G} = \frac{1}{\sqrt{2^{NM}}}\prod_{p} (\mathbb{I}+B_p)\ket{00\dots0},
\end{equation}
where $N\times M$ is the total number of plaquettes.
We note that the choice of $A_s$ and $B_p$ is dual to that originally used by Kitaev; both conventions are widely used in literature and can be related by a single layer of Hadamards. An
expansion of the product suggests $\ket{G}$ is an equal-weight linear combination of configurations with each plaquette $p$ acted on by $\mathbb{I}$ or $B_p$ with equal probability. This resembles an equal-weight superposition of all the binary digits with each binary representing the action of operators at $p$, with the relations $0\rightarrow \mathbb{I}$ and $1\rightarrow B_p = \prod_{i\in p} X_i$ (see Fig.~\ref{sm:fig:binary}). This motivates the following construction of toric code
\begin{enumerate}
    \item Initialize the product state $\ket{00...0}$ on all the qubits. 
    \item Identify a representative qubit for each plaquette.
    \item Apply Hadamard gate $H$ on each representative qubit.
    \item Within each plaquette, apply CNOTs controlled by the representative qubit and targeting the other qubits in the plaquette. Perform the control operation over all plaquettes in an order such that the state stored in the representative qubits are not changed until the CNOT operations in their plaquette have been applied. 
\end{enumerate}
Steps 1-3 initialize an equal weight superposition of all the binary strings of representative qubits, in Step 4 we apply the plaquette operator $ B_p$ on each plaquette depending on the representative qubit state, which turns a qubit binary string into a plaquette configuration. We illustrate this in further detail in Fig.~\ref{fig:sm_12q_illustration}.

To specify the state on a finite system, we need to impose boundary conditions.  
These can either be ``matching" or ``mixed," corresponding to whether the boundary plaquettes are of the same type or not, respectively.  
For the former, there is a unique ground state--we call this the toric code ground state in the main text.
For the latter, the ground state subspace is two-dimensional and can thus encode a logical qubit--we refer to these states as logical states of the surface code in the main text.
Other boundary conditions are possible, though we do not explore them here, such as periodic boundary conditions placing the code on a torus, or inserting defects inside the lattice.

\subsection{Matching boundary conditions}

\begin{figure*}[ht]
   \centering
   \includegraphics[width=504pt]{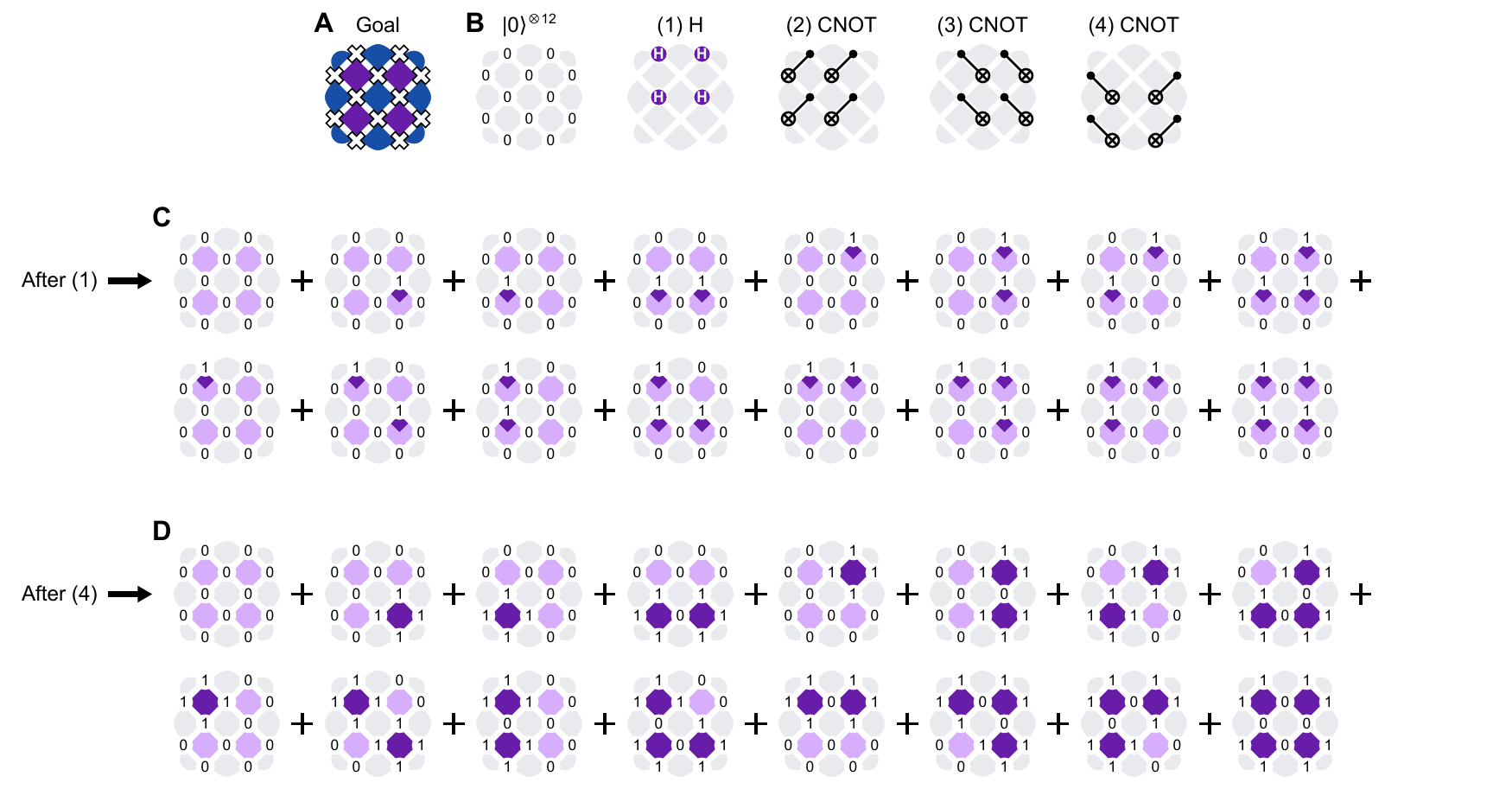}
   \caption{
   \textbf{State preparation illustration for 12 qubits, matching boundary.} 
   (\textbf{A}) Schematic showing the 12-qubit system with four plaquettes (purple), similar to Fig.~1A.
   (\textbf{B}) Quantum circuit to transform $\ket{0}^{\otimes 12}\to\ket{G}$, similar to Fig.~1B.
   (\textbf{C}) Wavefunction after the four Hadamard gates, a uniform superposition of $2^4=16$ bitstrings. Each Hadamard is associated with the plaquette (purple) below. We darken a portion of each plaquette underneath the 1's that came from its associated Hadamard. Each plaquette has a darkened portion in exactly half the bitstrings.
   (\textbf{D}) Wavefunction $\ket{G}$ after the complete circuit. 
   }
   \label{fig:sm_12q_illustration}
\end{figure*}

In the case of matching boundary conditions, the lattice consists of a rectangular array of complete plaquettes (see Figs.~1A and \ref{fig:sm_12q_illustration}). Following the design principle, all the qubits are initialized with state $\ket{0}$. We choose the top qubit on each plaquette to be the representative qubit.

To proceed, let us consider a system of two plaquette columns, shown in Fig~\ref{fig:sm_12q_illustration}. In panel B, we initialize the qubits, apply Hadamards on the representative qubits, and apply CNOTs from the representative qubits to the other qubits. Note that after B(1-3), the states of representative qubits are stored in the boundary qubits on the sides, so in panel B(4) the side qubits control the CNOTs, reducing the circuit depth.
Fig.~\ref{fig:sm_12q_illustration}C shows the wavefunction after the Hadamards.
The CNOTs act to ``spread out" the 1's to form loops around the plaquettes, effectively realizing $(\mathbb{I}+B_p)$.
As discussed in the main text, each bitstring is an eigenstate of all the stars $A_s$ (blue), and the superposition of all 16 is an eigenstate of all the plaquettes $B_p$ (purple): $B_p\ket{G}$ maintains the same superposition.
The situation for the 31-qubit system in Fig.~1A is analogous but intractable to draw, involving superpositions of $2^{12}=4096$ bitstrings.

For systems with more columns, we can grow the toric code starting from the middle out, following a similar strategy. This is shown in Fig.~1B. This larger circuit begins similarly to Fig.~\ref{fig:sm_12q_illustration}B, for the central two columns, and then extends outward (overlapping CNOT layers where possible to reduce depth). This procedure generalizes easily to wider systems with linear depth scaling, independent of the height.

\subsection{Mixed boundary conditions (logical states)}

\begin{figure*}[ht]
   \centering
   \includegraphics[width=504pt]{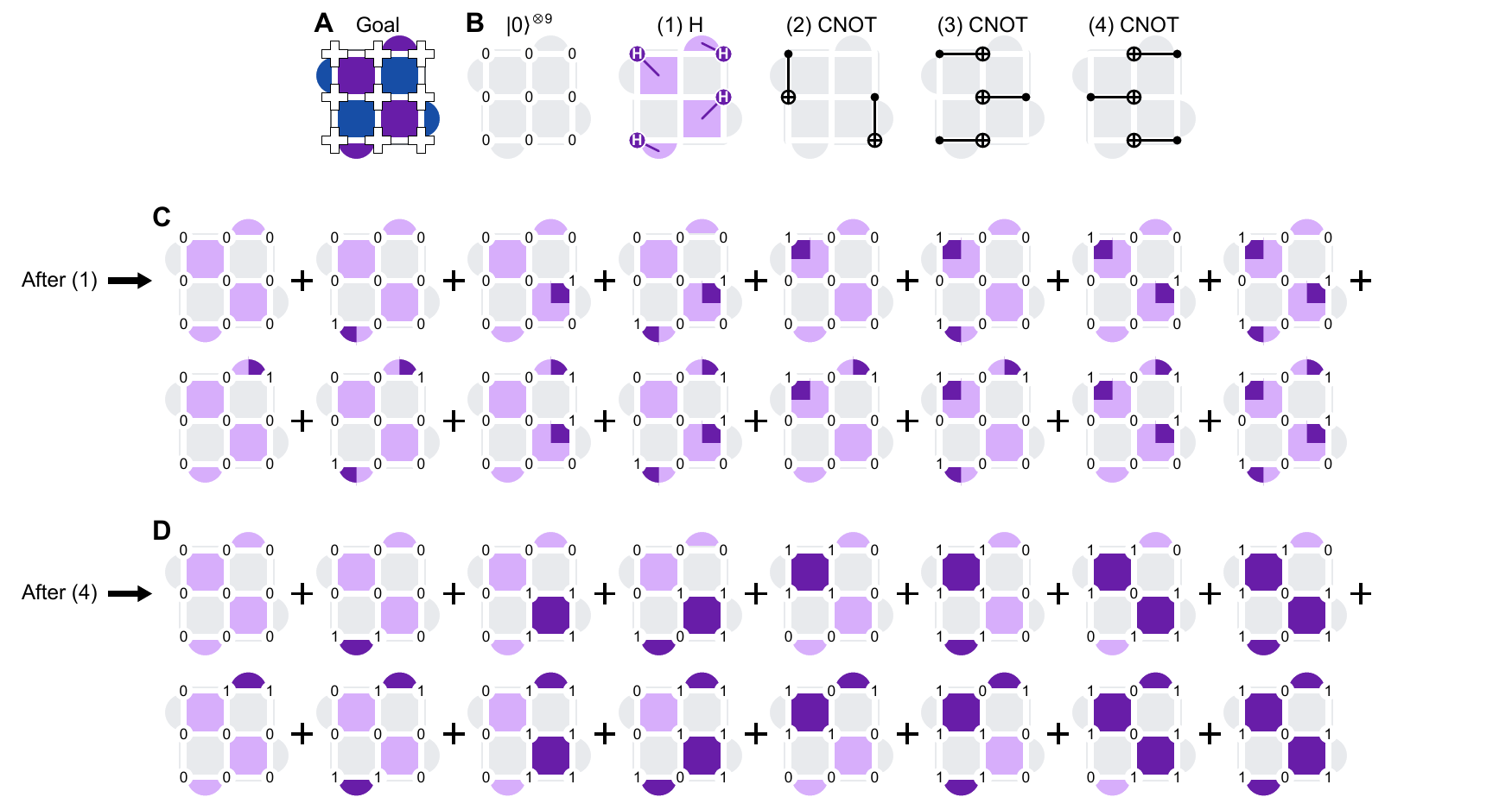}
   \caption{
   \textbf{State preparation illustration for 9 qubits, mixed boundary, $\ket{0_L}$.} 
   (\textbf{A}) Schematic showing a $3\times 3$ logical qubit with four plaquettes (purple), similar to Fig.~4A.
   (\textbf{B}) Quantum circuit to transform $\ket{0}^{\otimes 9}\to\ket{0_L}$. This maintains $Z_L=+1$.
   (\textbf{C}) Wavefunction after the four Hadamard gates, a uniform superposition of $2^4=16$ bitstrings. Each Hadamard is associated with a plaquette (purple). We darken a portion of each plaquette by the 1's that came from its associated Hadamard. Each plaquette has a darkened portion in exactly half the bitstrings.
   (\textbf{D}) Wavefunction $\ket{0_L}$ after the complete circuit. 
   }
   \label{fig:sm_9q_illustration}
\end{figure*}

\begin{figure*}[ht]
   \centering
   \includegraphics[width=504pt]{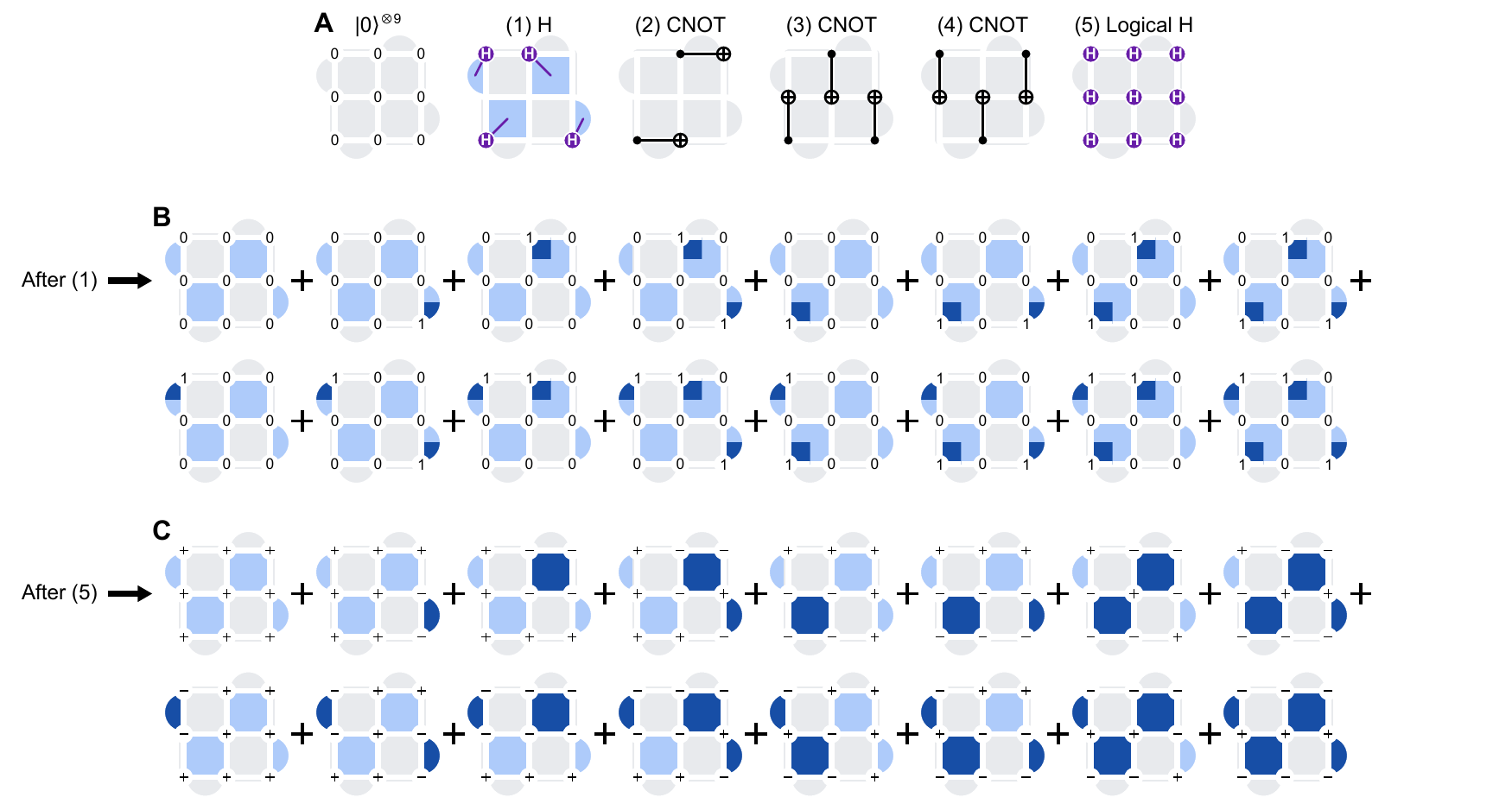}
   \caption{
   \textbf{State preparation illustration for 9 qubits, mixed boundary, $\ket{{+}_L}$.} 
   (\textbf{A}) Quantum circuit to transform $\ket{0}^{\otimes 9}\to\ket{{+}_L}$. Steps (1-4) are the same as Fig.~\ref{fig:sm_9q_illustration}B but rotated $90^\circ$. The final step is a transversal logical Hadamard \cite{Fowler2018}, which transforms $\ket{0_L}\to\ket{{+}_L}$ and effectively rotates the code $90^\circ$.
   (\textbf{B}) Wavefunction after the four Hadamard gates, a uniform superposition of $2^4=16$ bitstrings. Each Hadamard is associated with a star (blue). We darken a portion of each star by the 1's that came from its associated Hadamard. Each star has a darkened portion in exactly half the bitstrings.
   (\textbf{C}) Wavefunction $\ket{{+}_L}$ after the complete circuit. This is similar to $\ket{0_L}$ in Fig.~\ref{fig:sm_9q_illustration}D, but in $X$ basis and rotated by $90^\circ$. Each element in the sum is an eigenstate of all the plaquettes $B_p$, and the superposition of all 16 is an eigenstate of all the stars $A_s$ (blue). The state can also be written in $Z$ basis as $(\ket{0_L}+\ket{1_L})/\sqrt{2} = (\ket{0_L}+X_L\ket{0_L})/\sqrt{2}$.
   The situation for the $5\times 5$ system in Fig.~4A is analogous. To prepare $\ket{{+}_L}$ in the $5\times 5$ system, we rotate the circuit in Fig.~\ref{fig:sm_25q_circuits}A by $90^\circ$ and end with a transversal logical Hadamard.
   }
   \label{fig:sm_9q_plus_illustration}
\end{figure*}

\begin{figure*}[ht]
   \centering
   \includegraphics[width=504pt]{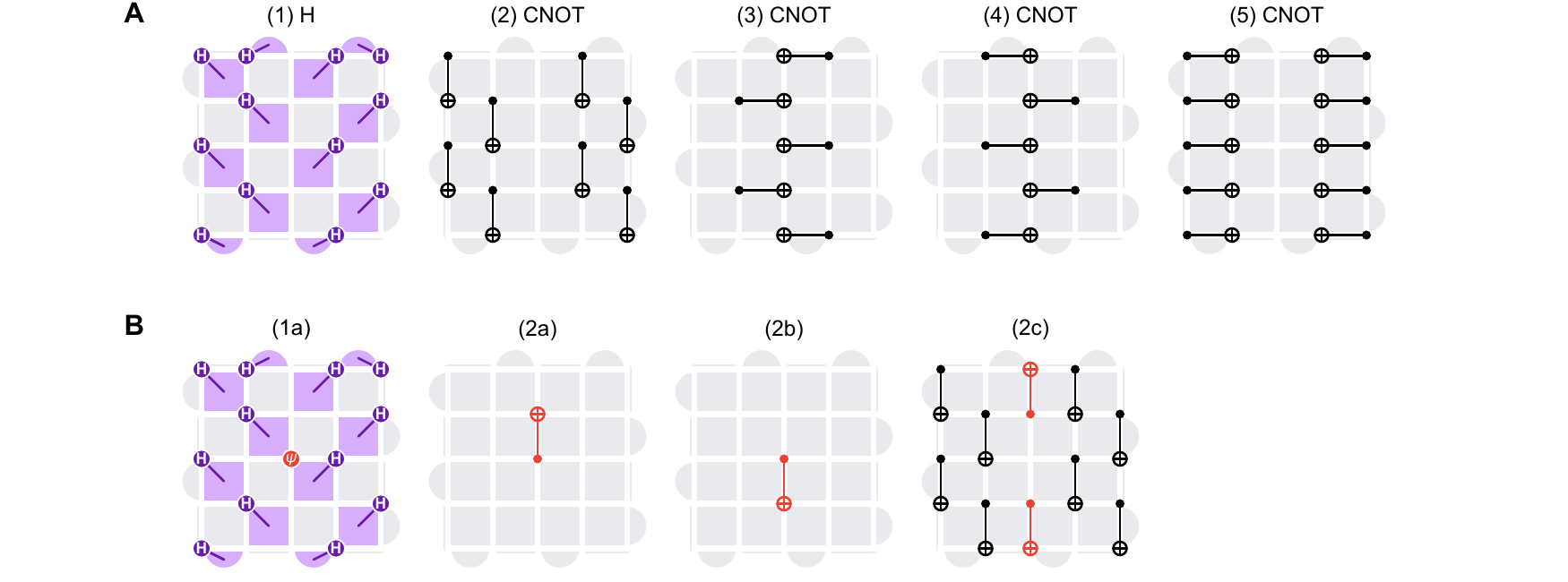}
   \caption{
   \textbf{State preparation and injection circuits for $5\times 5$ logical qubit states (mixed boundary).} 
   (\textbf{A}) Quantum circuit to transform $\ket{0}^{\otimes 25}\to\ket{0_L}$, similar to Fig.~\ref{fig:sm_9q_illustration}A. This maintains $Z_L=+1$ (see Fig.~4A).
   To prepare $\ket{{+}_L}$, we rotate the circuit $90^\circ$ and perform a transversal logical Hadamard at the end, as in Fig.~\ref{fig:sm_9q_plus_illustration}.
   (\textbf{B}) To inject an arbitrary logical state $\alpha\ket{0_L}+\beta\ket{1_L}$, we replace steps (1) and (2) from \textbf{A}, initializing the center qubit to the desired $\ket{\psi}=\alpha\ket{0}+\beta\ket{1}$.
   }
   \label{fig:sm_25q_circuits}
\end{figure*}

Mixed boundary conditions result in a two-dimensional ground state subspace that can encode a logical qubit.
The distance-3 and 5 surface code can be encoded on the lattice shown in Fig.~4A, where the plaquette (purple) and the star (blue) stabilizers correspond to $\prod_{i\in p}X_i$ and $\prod_{i\in s}Z_i$. On the boundary, some stars and plaquettes are incomplete, the stabilizers there are taken to be the product of Pauli operators on the two remaining bonds.

To construct the circuit for the distance-3 and 5 codes, we again follow the design principle above. The representative qubit is chosen as the outer-top qubit referenced to the center of the system. In the case of the incomplete plaquettes on the top boundary, we choose the outer qubit to be representative (see Fig.~\ref{fig:sm_9q_illustration} and Fig.~\ref{fig:sm_25q_circuits}).

Fig.~\ref{fig:sm_9q_illustration} shows the circuit construction to encode the logical state $\ket{0_L}$ for a distance-3 code, with many parallels to Fig.~\ref{fig:sm_12q_illustration} to help visualize the state. Here again, the CNOTs act to ``spread out" the 1's to form loops around the plaquettes, effectively realizing $(\mathbb{I}+B_p)$.
This state is a $+1$ eigenstate of $Z_L$ (see Fig.~4A).
The situation for the $5\times 5$ system in Fig.~4A is analogous but intractable to draw, involving superpositions of $2^{12}=4096$ bitstrings.
We can readily create $\ket{1_L}$ using $X_L\ket{0_L}$, where $X_L$ is simply a produce single-qubit $X$ gates.

To create $X_L$ eigenstates, we take advantage of the transversal logical Hadamard, where applying H to all the qubits performs a logical H and, as a side effect, also rotates the code $90^\circ$ \cite{Fowler2018}. To compensate, we simply rotate the $\ket{0_L}$ circuit $90^\circ$ and add the transversal Hadamard, as shown in Fig.~\ref{fig:sm_9q_plus_illustration}. This creates $\ket{{+}_L}$, and we can also readily create $\ket{{-}_L}=Z_L\ket{{+}_L}$.

These circuits generalize readily to larger circuits, such as the distance-5 case shown in Fig.~\ref{fig:sm_25q_circuits}A. Distance-$d$ requires only $(d+3)/2$ CNOT layers (for odd $d$).

By altering the beginning of the circuit, we can inject an arbitrary logical state. This is shown in Fig.~\ref{fig:sm_25q_circuits}B for the distance-5 case (it generalizes easily with depth linear in distance). The center qubit (red) is prepared in arbitrary single-qubit state $\alpha\ket{0}+\beta\ket{1}$. 
In (1a), we initialize the center qubit in $\ket{\psi}=\alpha\ket{0}+\beta\ket{1}$ along with the Hadamards. In (2a-c), we ``spread" this state along the qubits of $X_L$ (the five qubits in the center column; see Fig.~4A) using the CNOTs highlighted in red. This creates a GHZ-like state on those five qubits, $(\alpha\mathbb{I}+\beta X_L)\ket{00000}=\alpha\ket{00000}+\beta\ket{11111}$.
Step (2c) includes the final layer of red CNOTs as well as step (2) from Fig.~\ref{fig:sm_25q_circuits}A to minimize circuit depth. 
We then proceed with steps (3-5) from Fig.~\ref{fig:sm_25q_circuits}A.

\subsection{Circuit compilation and optimization}

\begin{figure}
    \centering
    \includegraphics{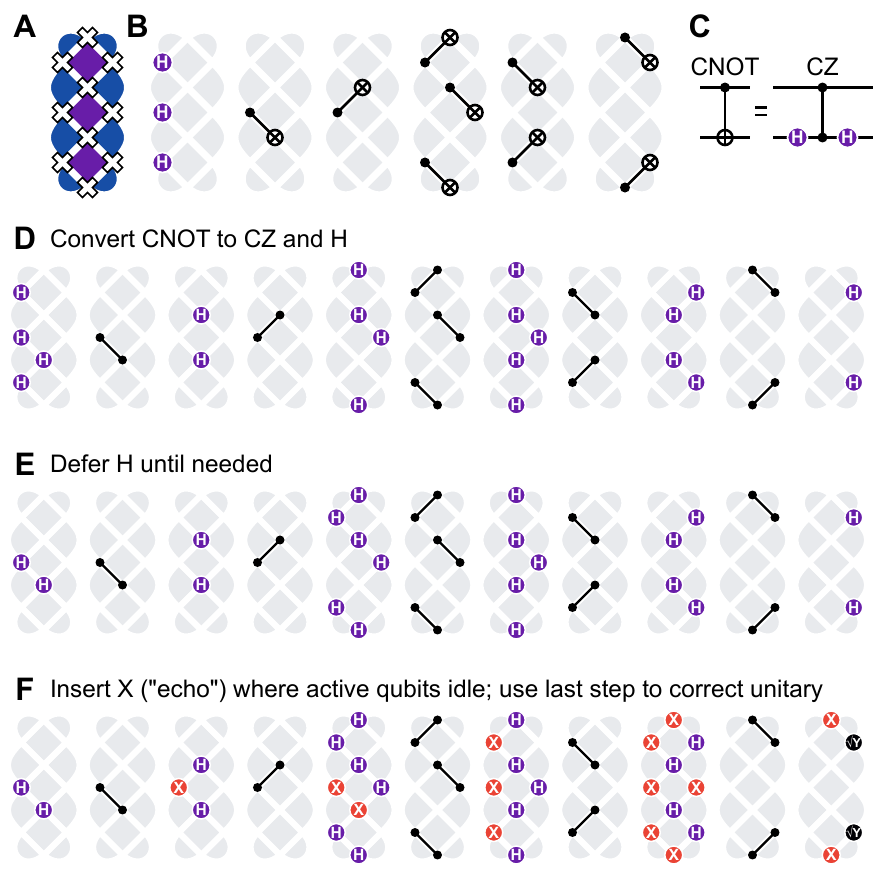}
    \caption{
    \textbf{Circuit compilation example.}
    (\textbf{A}) Example 10-qubit system with matching boundaries, similar to Fig.~\ref{fig:sm_12q_illustration}A.
    (\textbf{B}) Circuit to prepare $\ket{G}$, similar to Fig.~1B but rotated $90^\circ$. We intentionally use this orientation with 5 CNOT layers to illustrate the optimization steps.
    (\textbf{C}) Decomposition of CNOT into CZ and Hadamard.
    (\textbf{D}) Using \textbf{C}, convert \textbf{B} into CZ and H, preserving the CNOT layer structure.
    (\textbf{E}) Defer H gates to keep qubits longer in $\ket{0}$.
    (\textbf{F}) Insert $X$ gates to echo low-frequency noise. Once a qubit leaves $\ket{0}$, we do not let it idle between CZ layers. In the final step, we transform the single qubit gates to cancel out the effects of the inserted $X$ gates, in this case using $X$, $\sqrt{Y}$, and identity.
    }
    \label{sm:fig:sm_circuit_optimization}
\end{figure}

In Fig.~\ref{sm:fig:sm_circuit_optimization}, we walk through our circuit optimization techniques for an example state preparation circuit. We use these optimization steps on all of the circuits run in the main text except for the randomized compiling case (see Sec.~\ref{sm:subsection:randomized_compiling}).

%
%
\section{Readout error mitigation}\label{sm:readout}

\begin{figure}
    \centering
    \includegraphics{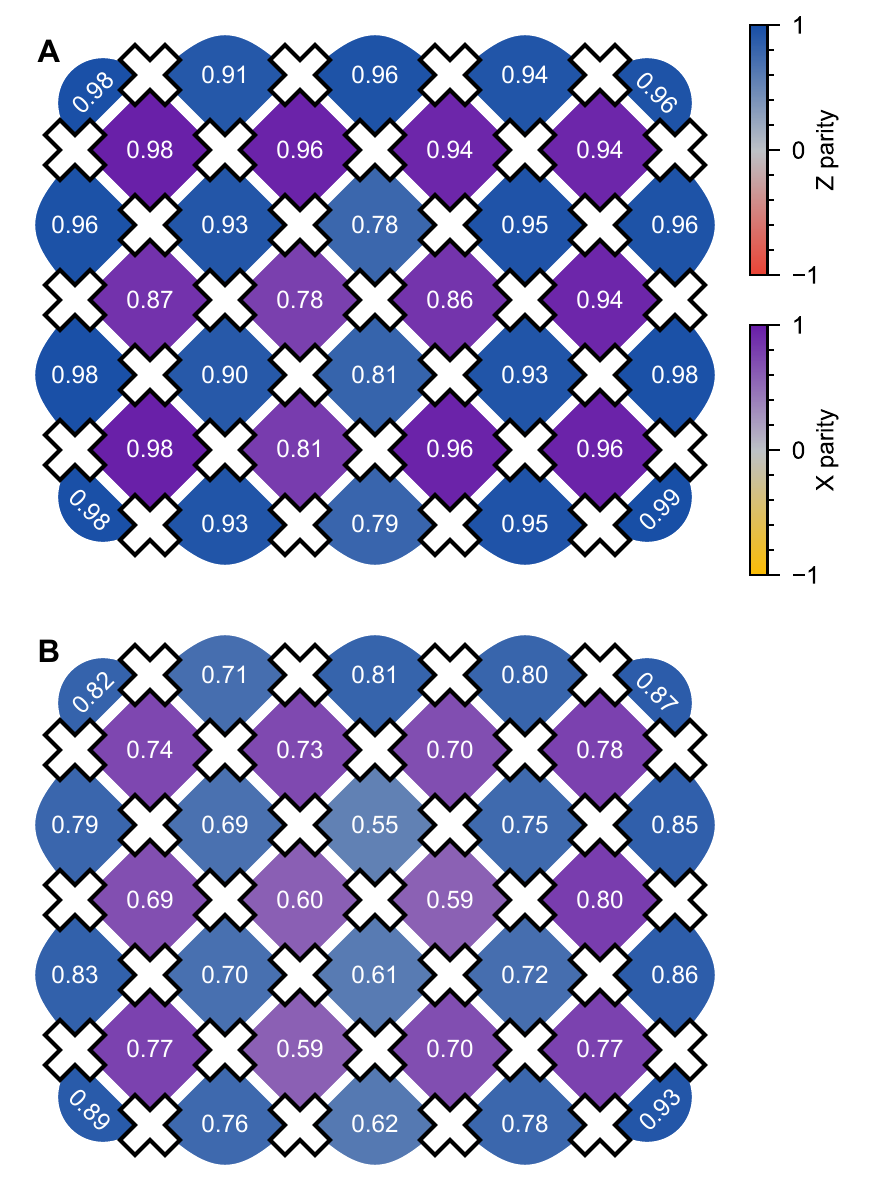}
    \caption{
    \textbf{Parity data with and without readout correction.}
    (\textbf{A}) Same as Fig.~1A. Before evaluating the parities, we correct each probability distribution using iterative Bayesian unfolding (see text).
    (\textbf{B}) Evaluating the parities directly from the measured probability distributions (no iterative Bayesian unfolding or other correction).
    }
    \label{sm:fig:sm_fig1a_unfolding}
\end{figure}

Measuring superconducting qubits is vulnerable to various errors, such as qubit decay, other unwanted qubit transitions, and separation error. 
Without full error correction, these readout errors severe limitations on the computational fidelity of quantum processors.
It is therefore important to mitigate the readout errors strategically when using NISQ devices.

Note we discuss and benchmark readout performance in Sec.~\ref{sm:benchmarks}, including discussion of related errors as state preparation and gate error, which we neglect here, since measurement error is dominant on this device.

One way to mitigate readout errors is using the response matrix~\cite{Nachman2020}. Suppose, for bitstrings $s,s'$, the observed probabilities are $P_o(s)$ and the actual error-free probabilities are $P_a(s')$.
This method assumes the two probability distributions are related by a response matrix $P(s|s')$ via $P_o(s) = \sum_{s'}P(s|s')P_a(s')$. In the experiment, the response matrix is obtained by a set of calibration experiments over computational basis states~\cite{Nachman2020}. This is done by preparing the product state $\ket{s}$ for some bitstring $s$, then measuring the probability distribution by repeated bitstring readouts. Such measurements are carried out for all the possible bitstring $s$. The measured probabilities are then used to approximate the response matrix. In this work, the number of repetitions used in each bitstring basis is 10000, 64000 and 64000 for 4-, 6- and 9-qubit error-mitigation, respectively.

The task of error-mitigation becomes a matter of inverting the response matrix to infer the actual distribution from the observed distribution; this procedure is known as ``unfolding" in high energy physics.
The unfolding of the response matrix can be performed with different methods. Here we employ iterative Bayesian unfolding (IBU)~\cite{Agostini1994}, where the unfolded distribution is inferred by recursively calling Bayes' theorem. This error-mitigation scheme is used to mitigate the readout errors in the parity measurements (see Fig.~\ref{sm:fig:sm_fig1a_unfolding}) and the entropy measurements (see Sec.~\ref{sm:rm}). The regularization parameter for IBU is the number of iteration steps, which can be chosen in advance, seeking optimal convergence.

The full response matrix is known to capture typical uncorrelated and correlated noise. However, obtaining and unfolding the full response matrix is in general exponentially costly for large systems. This limits the scalability of such error-mitigation techniques. For certain types of error models that are typical in the current superconducting qubits device (such as the uncorrelated errors), the cost of the calibration and the classical processing can be greatly reduced. This allows a possible scalable protocol for mitigating readout errors \cite{Bravyi2020}. 

\begin{figure*}[t]
  \centering
  \includegraphics{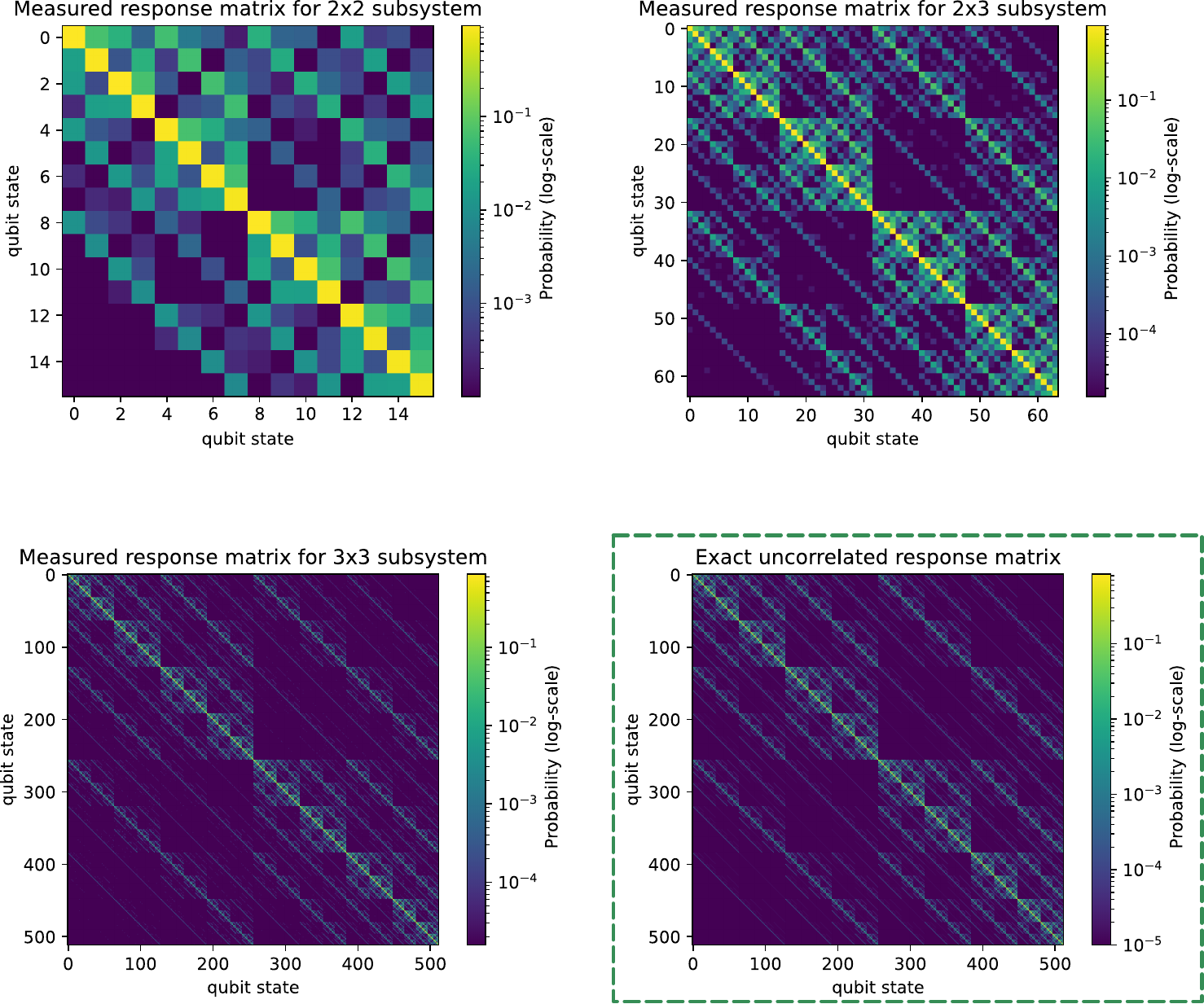}
  \caption{ {\bf Selected response matrices displayed in log-scale to highlight uncorrelated pattern.} Visualization of typical response matrices obtained in the error-mitigation calibration experiments. The colors are displayed in log-scale to highlight the uncorrelated noise pattern in the response matrices. We extract effective uncorrelated error rates from the matrices given respectively by ($2\times 2$) $e_0 = 0.016$ and $e_1 =0.056$, ($2\times 3$) $e_0 = 0.019$ and $ e_1 = 0.050$, ($3\times 3$) $e_00.018$ and $ e_1 = 0.048$. For comparison we show the image for an exact response matrix of $3\times 3$ subsystem for uncorrelated noise with error rates $e_0 = 0.02$ and $e_1 = 0.05$ (green dashed box).}
  \label{sm:fig:readout_matrix}
\end{figure*}

In our experiment, the structures of the response matrices were consistent with uncorrelated errors (see Fig.~\ref{sm:fig:readout_matrix} for typical response matrices measured in the experiments). This allows us to understand the error effects by Monte Carlo simulation. 

%
%
\section{Measuring topological entanglement entropy}\label{sm:rm}

Measuring the entropy of a system is experimentally challenging: one often needs the density matrix $\rho$, from which one can extract the von Neumann entropy 
\begin{align}
    S &= - \text{Tr} \left[ \rho \ln \rho\right],
\end{align}
or $n$-th order R\'enyi entropy 
\begin{align}
    S^{(n)} &= \frac{1}{1-n} \ln \left( \Tr\rho^n\right).
\end{align}

The entropy cannot be measured directly, but can be accessed through quantum state tomography of the density matrix.  
Full quantum state tomography is resource intensive, with cost typically scaling exponentially with the subsystem size.
Moreover, tomography produces a biased estimator~\cite{Schwemmer2015}, which can sometimes be tricky to account for.

The topological entanglement entropy is defined using von Neumann entanglement entropies for the subsystems~\cite{Kitaev06,Levin2006}.
In the case of Abelian topological order (such as the toric code), the same equation holds when the von Neumann entropies are replaced by second R\'enyi entropies~\cite{Flammia09,Bonderson17}.
This equivalence is helpful when investigating larger system sizes, as we can extract the second R\'enyi entropies from the statistical correlations of the subsystems using the technique of randomized measurement (RM)~\cite{vanEnk2012, Elben2018, Brydges2019}. A main advantage of this protocol is the direct access to the entropy without reconstructing the full state, significantly reducing the required number of measurements. It also provides a simpler way to remove the bias and understand the statistical errors for the estimation.
\subsection{Randomized measurement of second R\'enyi entropy \label{sm:sec:RMRenyi}}

In this work, we focus on the randomized measurement (RM) protocol that measures the second Rényi entropy using single-qubit random unitary. Consider a subsystem $A$, whose purity is given by
\begin{equation}
    \Tr(\rho_A^2) = 2^{N_A}\sum_{s,s'}(-2)^{-H(s,s')}\overline{P(s)P(s')},
    \label{sm:eq:rm}
\end{equation}
where $N_A, \rho_A$ is the number of qubits and the density matrix of $A$. The average is over the tensor product of single-qubit random unitaries which act on the qubits in $A$ and are independently drawn from the circular unitary ensemble (CUE). $s,s'$ are the binary strings in the computational basis with $H(s,s')$ outputting the hamming distance between them, and $P(s)$ denotes the probability of observing $s$. The second Rényi entropy is given by $S^{(2)}(\rho_A) = -\ln(\Tr(\rho_A^2))$. A nice feature of the randomized method is that the same set of measurement data can be used to compute the entropies for multiple subsystems at the same time. This renders particular convenience in measuring the $S_\text{topo}$, which is inferred from a linear combination of the entropies from different partitions. In the experiment, we only have to measure the entropy of the subsystems themselves, from which $S_\text{topo}$ can be obtained by calculating all the entropies for different partition using the same data. This avoids having several randomized measurements on the subsystem partitions and the large statistical errors built up from adding and subtracting these independently-measured entropies.

In practice, $P(s)^2$ is a biased estimator for $\mathbb{E}(P(s))^2$ and needs to be replaced with an unbiased estimator 
\begin{equation}
    P\rightarrow P\times\frac{nP-1}{n-1},
\end{equation}
where $n$ is the number of measurements used to determine $P(s)$ \cite{Vermersch2018}. The random unitaries can be drawn from the continuous (Haar) measure. However, on many current devices it is more desirable to use a given finite set of pre-calibrated quantum gates. This is made possible by approximating the ensemble (up to certain statistical moment) using a unitary 3-design, e.g. the Clifford group \cite{zak2016,zhu2017}. The single-qubit random unitary can be implemented as random single-qubit Clifford gates. 

In the setting of RM, averaging over the tensor product of single-qubit Clifford gates is equivalent to averaging over all the Pauli basis measurement \cite{Huang2020}. This can be seen by noting the qubit measurement projects the state onto Pauli $Z$ basis, i.e. $\dyad{0} = (1+Z)/2$ and $\dyad{1} = (1-Z)/2$. The single-qubit Clifford gates send the Pauli $Z$ to any other non-identity Pauli gates with equal frequency, $U^{\dag}ZU = \pm P$, where $U\in \text{Cliff}(2)$ and $P\in\{X,Y,Z\}$. One can then go back to the Pauli $Z$ basis as in the usual Pauli measurements. The mapping with phase -1 corresponds to a bit-flip when transforming back to the Pauli $Z$ basis. The Hamming prefactor in Eq.~\eqref{sm:eq:rm} is preserved under bit-flip on $s,s'$, hence the equivalence follows.

\begin{figure}
    \centering
    \includegraphics{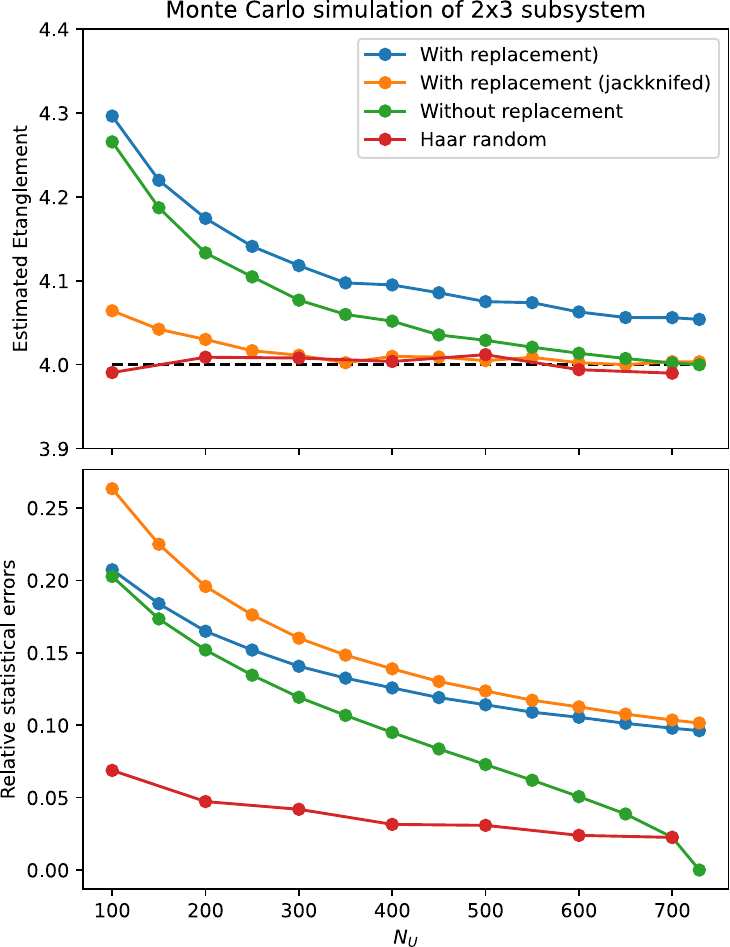}
    \caption{{\bf Monte Carlo simulation of entropy estimation.} We show the difference of using the discrete random Pauli rotations and using Haar-random unitary for estimating the second R\'enyi entropy of a 6-qubit subsystem ($2\times 3$ qubit array) in toric code. In the simulation, the bitstring probability distributions are measured perfectly in each random unitary instance. We use 200 Monte Carlo experiments for the Haar-random case and $>10^4$ experiments for the discrete Pauli rotation case. The Haar-random unitary protocol shows much smaller statistical errors and bias in the estimation.}
    \label{sm:fig:PaulivsHaar}
\end{figure}

Despite the properties of unitary 3-design, discretizing the random unitary measure with the Clifford group can give rise to different behaviour in the statistical fluctuations of the entropy measurement. In the example of toric code subsystems, the statistical errors are observed to be much larger in the random Clifford/Pauli case than in the Haar-random case. We illustrate this with the Monte Carlo simulation in Figure.~\ref{sm:fig:PaulivsHaar}. In the simulation, we estimate the second Rényi entropy of the $2\times 3$ (6-qubit) subsystem within the toric code assuming the bitstring probabilities can be measured perfectly, and compare the results of using random Pauli basis rotations and using Haar-random unitaries. To draw the random Pauli rotations from a finite set of $3^6 = 729$ elements, we can either sample with or without replacement. For the entropy estimation, the random Pauli protocol is observed to be biased upward. In the case when the samples are statistically independent (sampling with replacement), the bias can be mitigated by a jackknife resampling technique~\cite{Tukey1958}. The Haar-random case, however, produces nice estimation with minimal bias. The relative statistical errors reveal the advantage of the Haar-random protocol when the number of random unitaries drawn $N_U$ are fewer than the full set of Pauli rotations, with much smaller statistical errors than the other protocols. When $N_U\geq 3^6$, random sampling becomes unnecessary, as we can simply sum over all the possible Pauli rotations to obtain the exact average over the ensemble, resulting in zero statistical errors.

This motivates the RM experiments to use the random Pauli rotations in 4- and 6-qubit systems where the measurements over the full Pauli basis are feasible ($3^4 = 81$ for 4-qubit and $3^6 = 729$ for 6-qubit, equivalent to full tomography), and use 1000 instances of Haar-random single-qubit unitaries in the 9-qubit system where a scan of the full Pauli basis is not feasible ($3^9= 19683$ basis states). For each instance of the single-qubit unitaries, we repeat the bitstring measurement 10000 times, and then we attempt to correct the measured probability distribution for readout error (see Sec.~\ref{sm:readout}). 
\subsection{Unbiased estimator under error-mitigation}
A subtlety arises when applying the error unfolding in RM of the entropy, which is estimated from a list of cross-probabilities $P(s)P(s')$ between two bit-strings $s,s'$. However, the multinomial nature indicates finite covariance between $P(s)$ and $P(s')$. In other words, the estimation of $P(s)P(s')$ becomes biased:
\begin{equation}
\mathbb{E}[P(s)P(s')]-\mathbb{E}[P(s)]\mathbb{E}[P(s')] = \text{cov}(P(s),P(s')).
\end{equation}
In practice, an unbiased estimator for the cross-probability is used to remove this bias (see Section~\ref{sm:sec:RMRenyi}). 
That only allows us to remove the bias in the estimation based on the observed data, but not the error-mitigated data after IBU. 
To fix this deficiency, we need to simultaneously unfold the covariance during the iterative steps~\cite{Adye2011} and use the unfolded covariance to remove the bias in the error-mitigated estimation at the end.

In the experiment, the unfolding and the propagation of the covariance were performed using RooUnfold package \cite{Adye2011}. 
The iterative steps are chosen to be 15, 50 and 50 for the 4-qubit, 6-qubit and 9-qubit systems.

\begin{figure}
    \centering
    \includegraphics{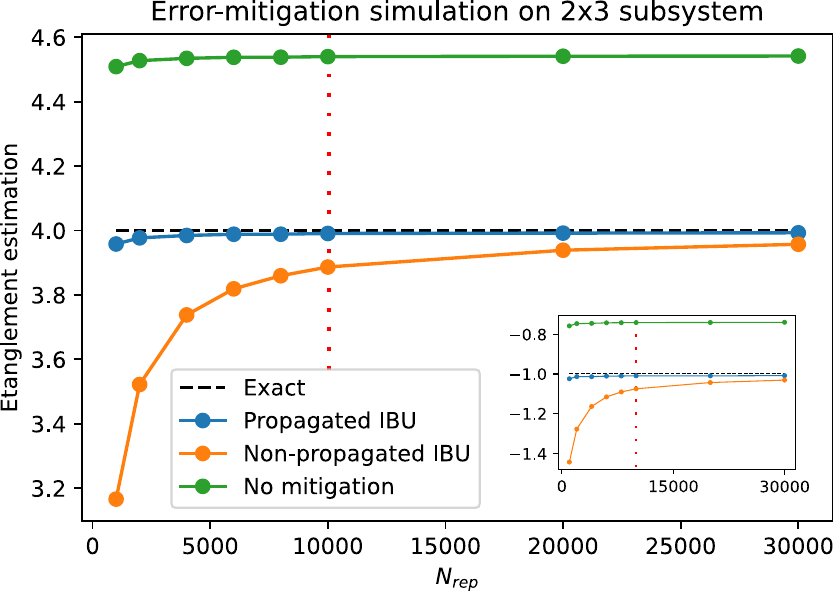}
    \caption{{\bf Monte Carlo simulation of the error-mitigation for a 6-qubit subsystem ($2\times 3$ qubit array) in toric code.} Error mitigation can be applied to estimate non-linear quantity like entropy, but the induced biased needs to be removed by error-propagation. Here we show a simulation of entanglement entropy estimation for a $2\times 3$ subsystem, using an uncorrelated noise model ($e_0 = 0.01$, $e_1 = 0.05$). The red dotted line highlights the number of repeated bitstring measurements used in the actual experiment. The inset shows a simulated estimation of $S_\text{topo}/\ln 2$.}
    \label{sm:fig:IBU_mitigation}
\end{figure}

To illustrate the effectiveness of the error-mitigation and the unfolding techniques, we simulate the 6-qubit experiment using an uncorrelated readout error model. The model assumes asymmetric error rates $e(0\rightarrow 1) = 0.01$ and $e(1\rightarrow 0) = 0.05$ during the readout procedure, while other sources of error such as decoherence and gate error are neglected. The randomized measurements are simulated by drawing sufficiently many single-qubit random unitaries (or equivalently, summing over the full Pauli basis rotations) and repeating the bitstring measurements $N_\text{rep}$ times for each random unitary instance, where $N_\text{rep}$ sweeps through a range of possible values. The results are shown in Figure.~\ref{sm:fig:IBU_mitigation}. Without taking error-propagation into account, a clear under-sampling bias is induced, where the estimated entropy will strongly depend on $N_\text{rep}$. Instead, when the error-propagation is taken into account, the bias is removed, allowing an accurate determination of the entropy for a wide range of bitstring samples $N_\text{rep}$.

\begin{figure*}[t]
  \centering
  \includegraphics{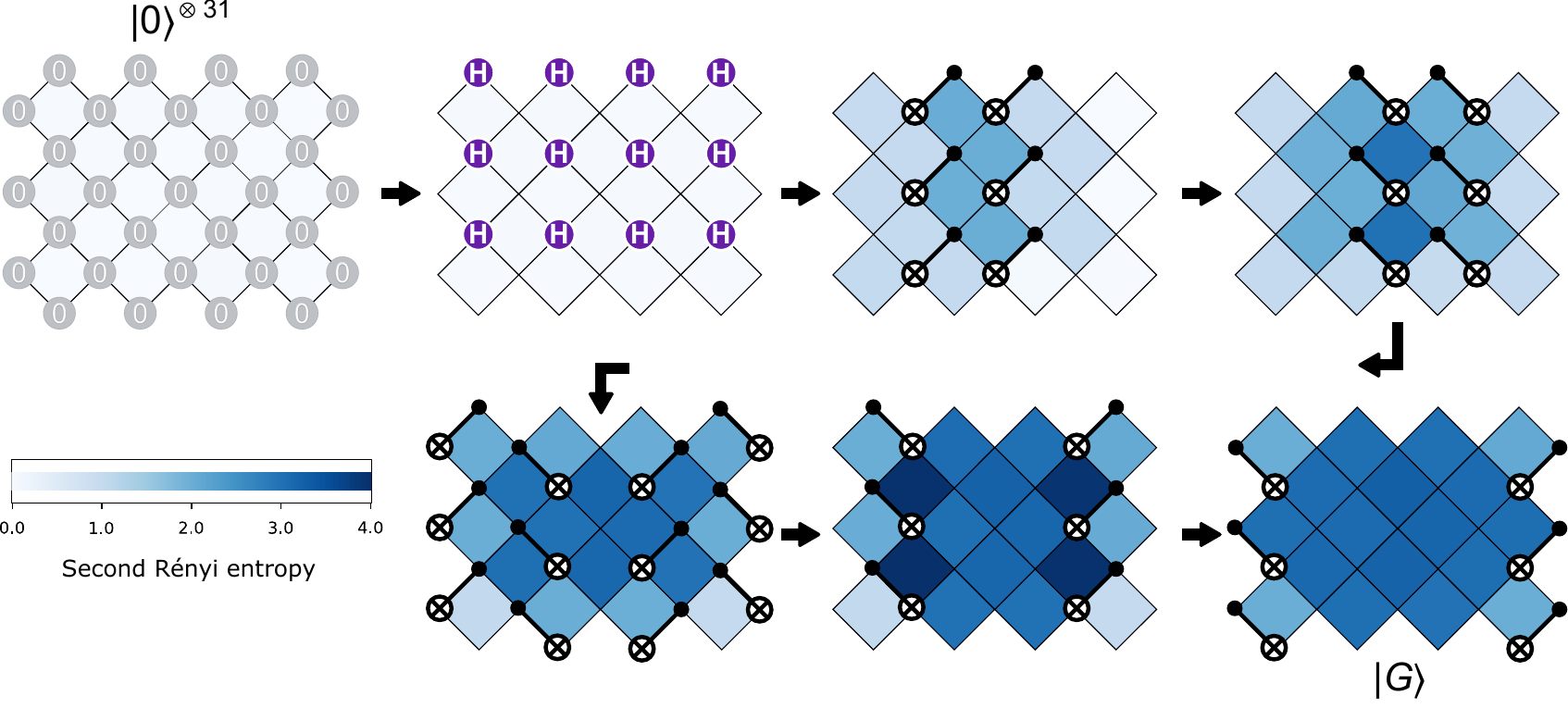}
  \caption{ {\bf Experimental snapshots of the second R\'enyi entropies during the ground state preparation steps.} We probed the second R\'enyi entropy of all the $2\times 2$ subsystems during the state preparation steps using randomized measurement, measuring over all $3^4=81$ Pauli basis combinations (10000 repetitions per basis). The values are shown in units of $\ln 2$ and  are consistent with the predicted values. Note that the corner plaquettes end with an entropy of 2, as expected due to the boundary conditions.}
  \label{sm:fig:sm_steps_entropy}
\end{figure*}

\subsection{Extended experimental details}
As a check for the state quality and the success of the entropy measurement protocol, we probed the (second R\'enyi) entropy for the 4-qubit subsystems by taking snapshots of the entropy values after each step of the state preparation in Fig.~1B. The entropies of all the 4-qubit subsystems at the stars and plaquettes are measured following the protocol  described above, giving a sequence of the 4-qubit subsystem entropies at each step as shown in Figure.~\ref{sm:fig:sm_steps_entropy}. The measured entropies closely match the ideal values (by carefully following the CNOT gates), demonstrating the quality of the state and entropy measurements.

\begin{figure}
    \centering
    \includegraphics{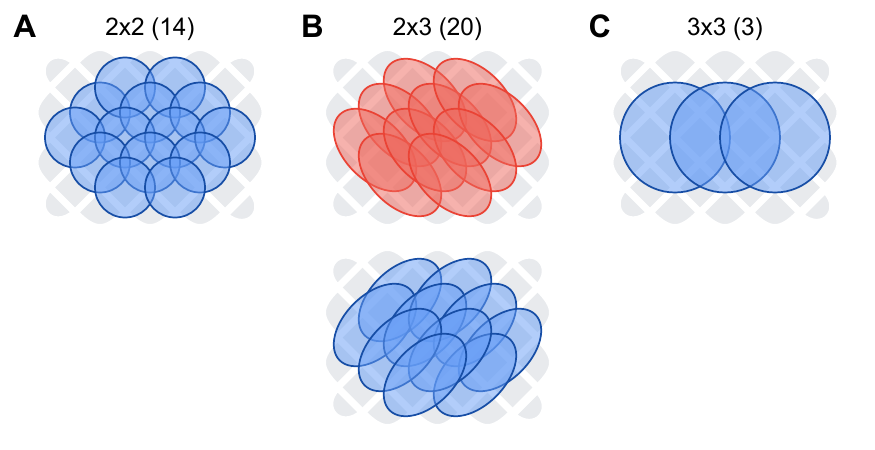}
    \caption{
    \textbf{Systems used for $S_\text{topo}$ measurements.}
    Refer to the $2\times 2$, $2\times 3$, and $3\times 3$ systems in Fig.~2A.
    As discussed in the main text, for the entropy data in Fig.~2C-D, we measure several of each system shape across the 31-qubit toric code ground state (see Fig.~1A).
    (\textbf{A}) $2\times 2$ systems (14). Note we exclude the corners which have different entropy; see Fig.~\ref{sm:fig:sm_steps_entropy}.
    (\textbf{B}) $2\times 3$ systems (20). For clarity, we split into two groups with different orientations.
    (\textbf{C}) $3\times 3$ systems (3).
    }
    \label{sm:fig:sm_entropy_systems}
\end{figure}

\begin{figure*}[t]
  \centering
  \includegraphics{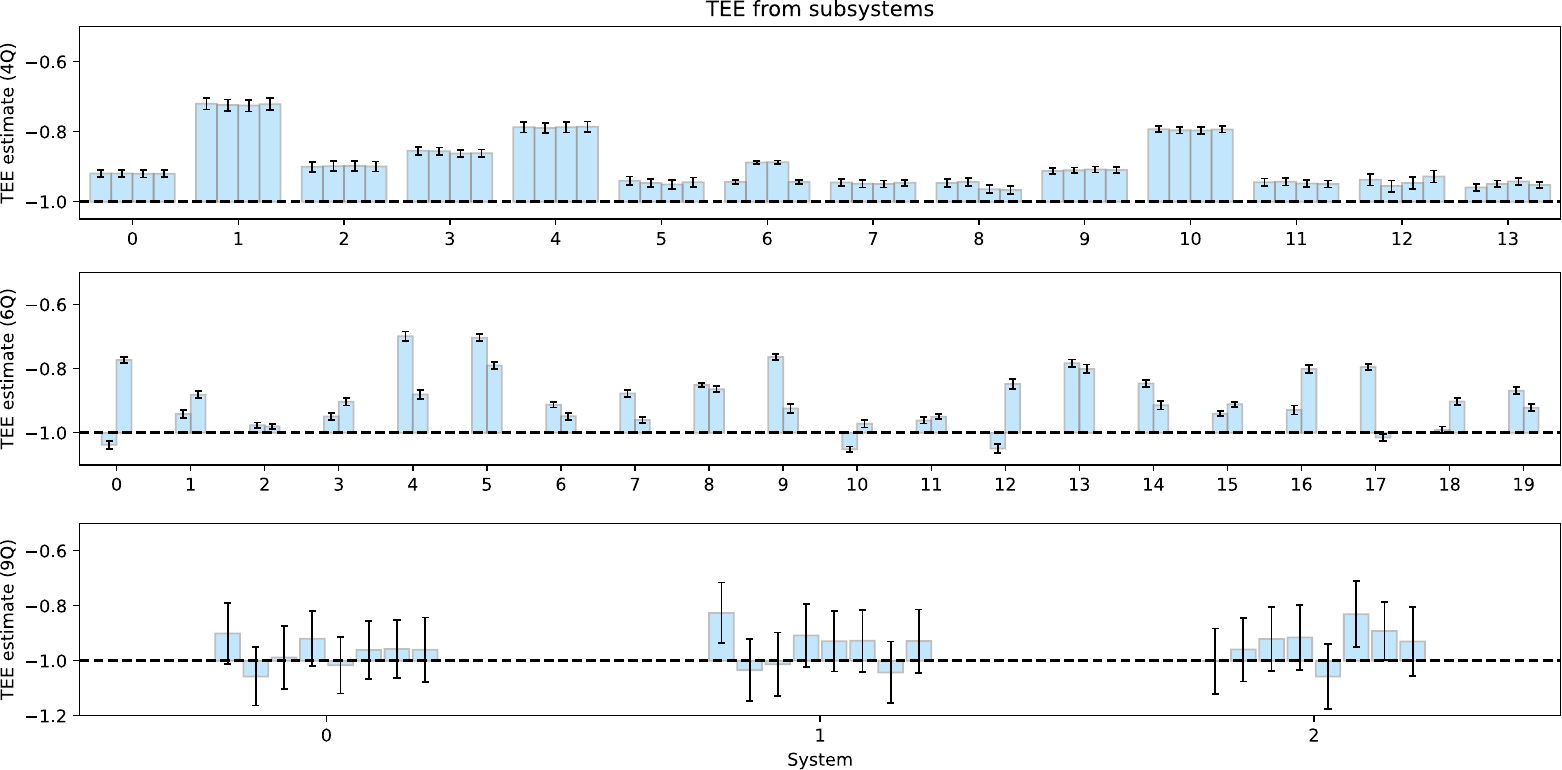}
  \caption{ {\bf The individual values of $S_\text{topo}$ of each subsystem for different partitions into subregions.} Here we present details for the histograms shown in Figure.~2D. In units of $\ln 2$, the expected topological entanglement entropy for the toric code is -1. The average relative statistical errors are $1.3\%$. $1.2\%$ and $12\%$ for the 4-, 6- and 9-qubit subsystems, respectively.}
  \label{sm:fig:uncertainty_subsys_topo}
\end{figure*}

As mentioned earlier, the topological entanglement entropy ($S_\text{topo}$) can be computed using a single set of randomized measurements data for a given subsystem. In our experiments, we perform the $S_\text{topo}$ measurements on 14 4-qubit subsystems, 20 6-qubit subsystems and 3 9-qubit subsystems across the device (see Figure.~\ref{sm:fig:sm_entropy_systems}). For each subsystem, we extracted multiple $S_\text{topo}$ values based on different partitions. By rotation and reflection, we can have 4, 2 and 8 ways to partition the 4-, 6- and 9-qubit subsystems. The $S_\text{topo}$ distributions for all these values are summarized in Fig.~2. In Fig.~\ref{sm:fig:uncertainty_subsys_topo}, we present the individual $S_\text{topo}$ values with estimated error bars of one standard deviation. We estimate the statistical errors with bootstrapping~\cite{EfroTibs93}. Despite the much larger Hilbert space, the 9-qubit subsystems show small uncertainty with an average relative statistical errors of $12\%$. This compares favorably against the uncorrelated error modelling on a 9-qubit subsystem with asymmetric error rates $e(0\rightarrow 1) = 0.01$ and $e(1\rightarrow 0) = 0.05$, estimating a relative statistical errors of $13\%$.

\begin{figure}
    \centering
    \includegraphics{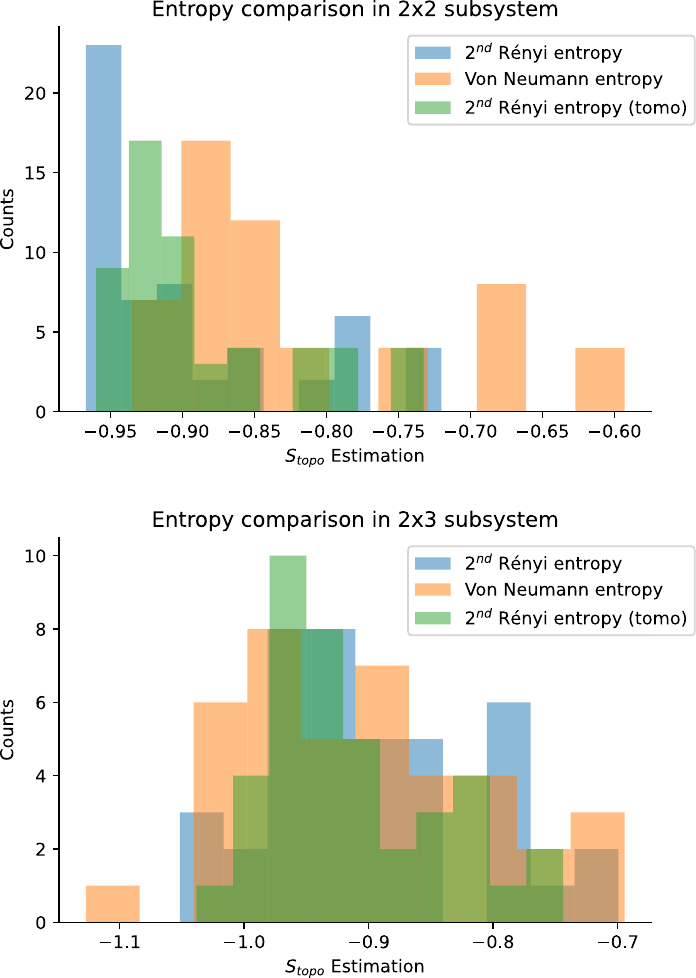}
    \caption{{\bf Comparison between the topological entanglement entropy estimation obtained with randomized measurements and quantum state tomography.} The full Pauli basis rotation data can be analyzed using randomized measurements and also quantum state tomography. We compare both cases to check consistency. 
    Top: $2\times 2$ subsystems. Distribution mean and standard deviation: $S^{(2)}_{\text{rand}} = -0.89\pm 0.07$, $S_{\text{tomo}} = -0.82\pm 0.1$ and $S^{(2)}_{\text{tomo}} = -0.89\pm 0.06$. 
    Bottom: $2\times 3$ subsystems, $S^{(2)}_{\text{rand}} = -0.90\pm 0.09$, $S_{\text{tomo}} = -0.91\pm 0.1$ and $S^{(2)}_{\text{tomo}} = -0.91\pm 0.07$.}
    \label{sm:fig:entropy_compare}
\end{figure}

The data for the 4- and 6-qubit randomized measurements were taken by measuring the bitstring probability distributions over all the Pauli bases, which is equivalent to a full quantum state tomography. We can therefore analyze the same set of data with standard quantum state tomography techniques. We find the reduced density matrices using maximum likelihood estimation through convex optimization. In Figure.~\ref{sm:fig:entropy_compare}, we show the comparison of the estimated $S_\text{topo}$ between the tomographic analysis and the randomized analysis on the same sets of data. The direct access to the density matrices also allows the computation of the corresponding von Neumann entropy for the subsystem, which is the usual entropy measure used to define $S_\text{topo}$, as opposed to the second R\'enyi entropy. The consistency between the $S_\text{topo}$ values obtained through different analysis of the same data again supports the reliability of the experiments.

\begin{figure*}[t]
  \centering
  \includegraphics{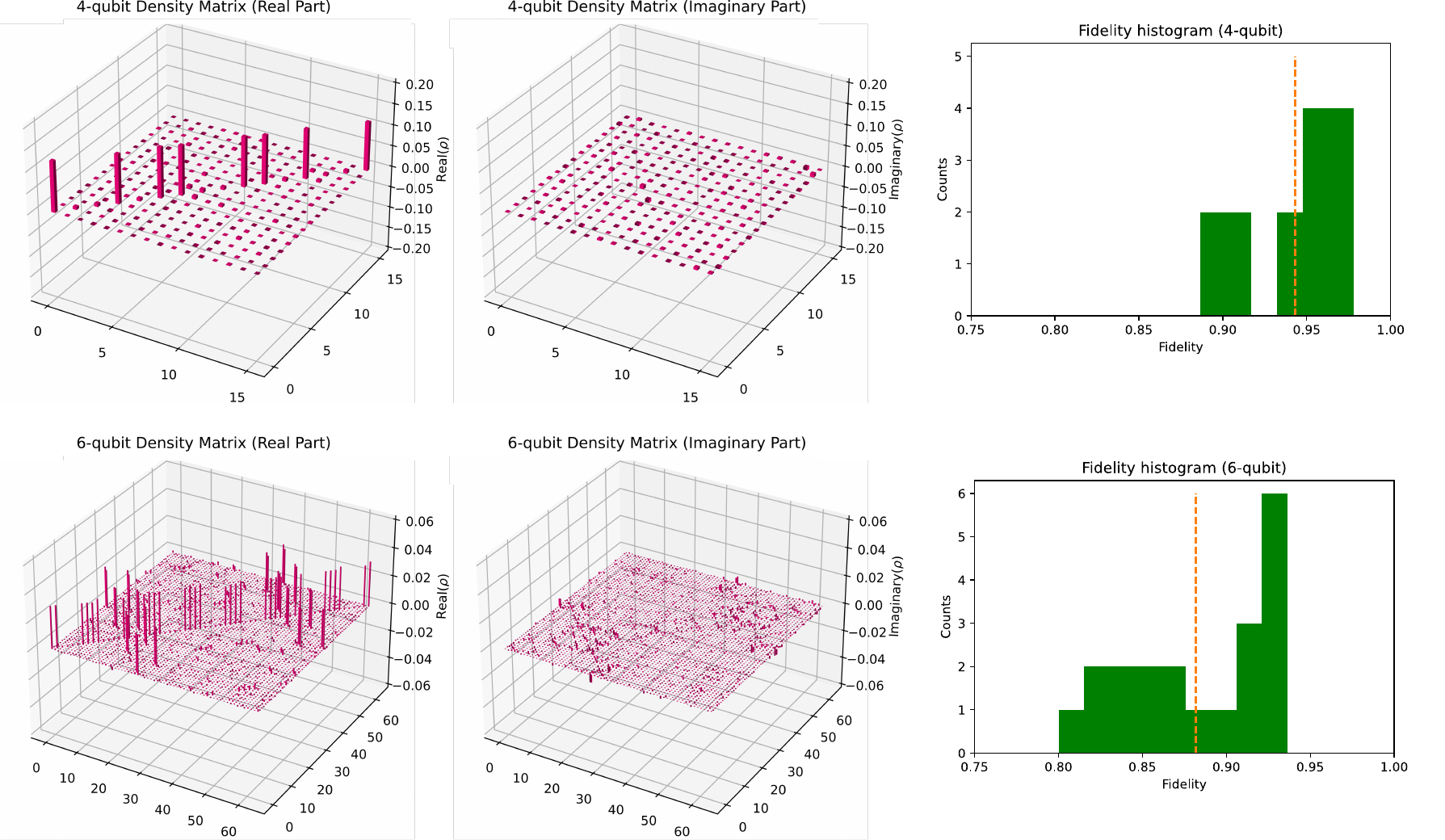}
  \caption{ {\bf Example density matrices and fidelity histogram.} The left panels show examples of 4- and 6-qubit measured density matrices. The density matrices are obtained using maximum likelihood estimation based on full quantum state tomography (with 10000 repeated measurements in each basis). The right panels show the histograms of fidelity for all the measured density matrices against the corresponding toric code subsystems. The average fidelity reaches $94\%$ and $88\%$ for 4- and 6-qubit states, respectively.  On average, we estimate the uncertainty of the fidelity to be 0.004 and 0.002 for 4- and 6-qubit subsystems, respectively. The shown examples are chosen to be near the average, having fidelity $94\%$ (4-qubit) and $91\%$ (6-qubit). }
  \label{sm:fig:tomo}
\end{figure*}

In Figure.~\ref{sm:fig:tomo}, we show the examples of reduced density matrices for 4- and 6-qubit subsystems obtained by tomographic analysis. The non-trivial entanglement pattern of the states are manifested by the rank deficiencies in the matrices. We can further extract the state fidelity of the subsystems against the exact toric code, which is summarized in the histograms in Figure.~\ref{sm:fig:tomo}. The average fidelity of the 4- and 6-qubit subsystems have reached $94\%$ and $88\%$ respectively.

%
%
\section{Simulating braiding}

In this Appendix, we elaborate on the protocol for simulating braiding in Fig.~3 of the main text, \textit{cf.} Fig. \ref{fig:sm_fig3_decompositions}.  
Exchange statistics refers to interchanging the position of a pair of identical anyons, while mutual statistics refers to exchanging the positions of two (possibly distinct) anyons twice.  
Equivalently, mutual statistics arise when circling one anyon around another (the two pictures are related by switching to the reference frame of one of the anyons).  
While fundamental particles have trivial mutual statistics (all +1) and exchange statistics ($+1$ for bosons and $-1$ for fermions), braiding Abelian anyons can result in more general phases.

\subsection{Interferometry}

The interferometric protocol is motivated by the simple quantum optics picture in which a single light source is split into two paths that interfere when recombined.  
In our digital quantum processor, we use an auxiliary qubit $a$ which is initially prepared as $ (\ket{0} + \ket{1})/\sqrt{2}$ to ``split" the target state $\ket{\varphi}$ into a superposition
\begin{equation}
    \ket{\Psi} = \ket{0}\otimes\ket{\varphi}+\ket{1}\otimes U\ket{\varphi}
\end{equation}
by an controlled-$U$ operation using the auxiliary $a$. Then 
\begin{equation}
    \ev{U}{\varphi} = \ev{X_a}{\Psi} + i\ev{Y_a}{\Psi},
\end{equation}
where $X_a,Y_a$ are single-qubit Pauli operators that act on $a$.
This procedure thus allows an experimental measurement of the overlap $\ev{U}{\varphi}$ for some state $\ket{\varphi}$ and local unitary $U$. The final overlap can be measured by single-qubit tomography of the auxiliary qubit $a$.

\begin{figure}
   \centering
   \includegraphics[width=252pt]{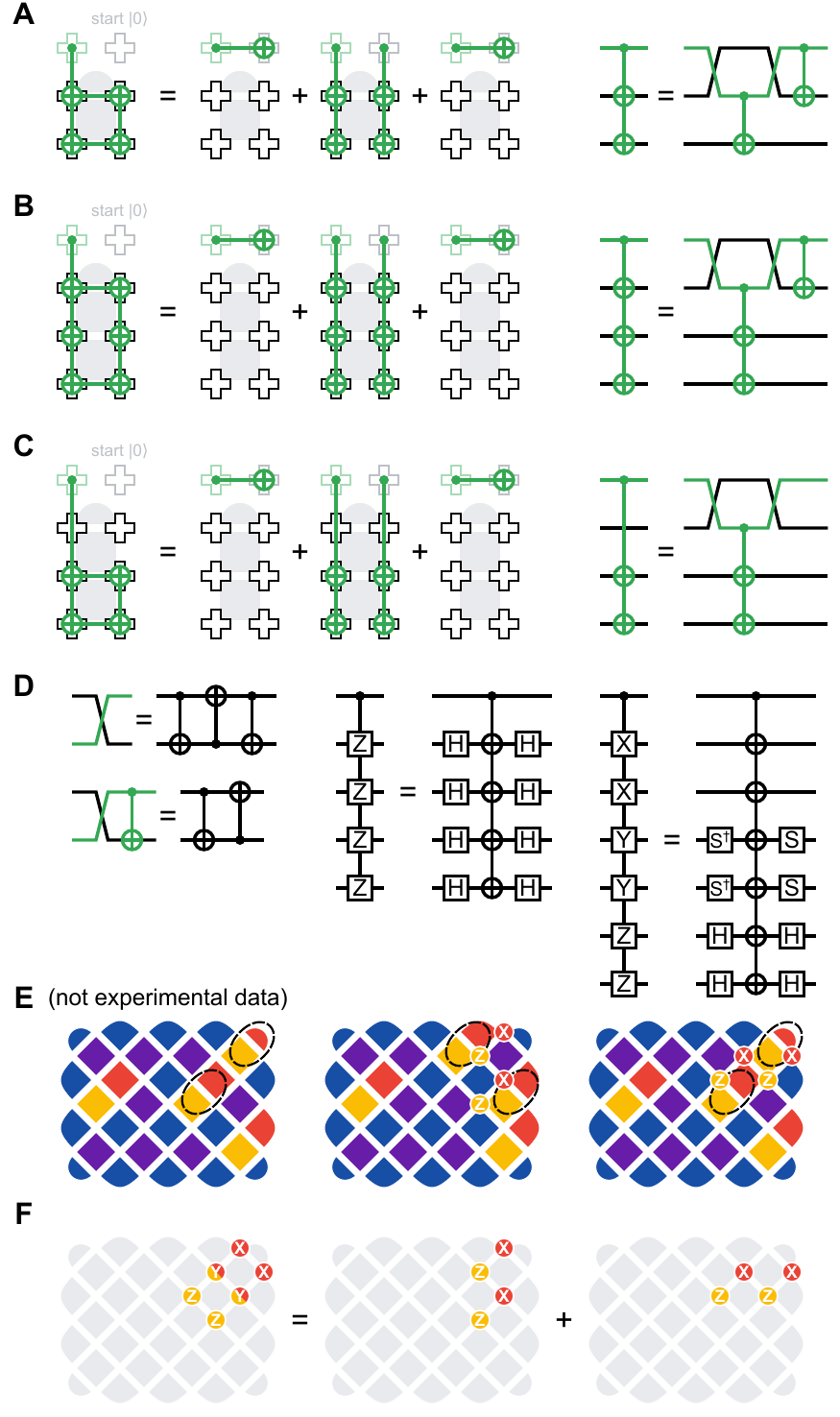}
   \caption{
   \textbf{Circuit decompositions for Fig.~3.} 
   (\textbf{A}) Circuit decomposition of controlled-\textit{XXXX} for Fig.~3D. We use a second auxiliary qubit, initially in $\ket{0}$, to decrease the circuit depth. We decompose into swap and CNOT (see \textbf{D} for further decomposition).
   (\textbf{B}) Circuit decomposition of controlled-\textit{XXXXXX}, which is used for Fig.~3E.
   (\textbf{C}) Circuit decomposition of controlled-\textit{XXXX} targeting qubits deeper in the array, which is used for other interferometry experiments.
   (\textbf{D}) Additional circuit decomposition details. Left: Conversions between swap and CNOT. Middle and right: example conversions between controlled operators using single-qubit rotations (S is $Z^{1/2}$). Ultimately, everything is compiled into CZ gates, and we use randomized compiling on these circuits when we extract the phases (see Sec.~\ref{sm:subsection:randomized_compiling}).
   (\textbf{E}) Schematic showing the idea behind the controlled-\textit{XXYYZZ} used in Fig.~3E. In two steps, we exchange the location of two $\psi$'s. Unlike the similar plots in the main text, these are not experimentally-measured parities. 
   (\textbf{F}) Simplification of the two-step sequence in \textbf{E} to a single step \textit{XXYYZZ}, as used in Fig.~3E.
   }
   \label{fig:sm_fig3_decompositions}
\end{figure}

\begin{figure*}
   \centering
   \includegraphics[width=504pt]{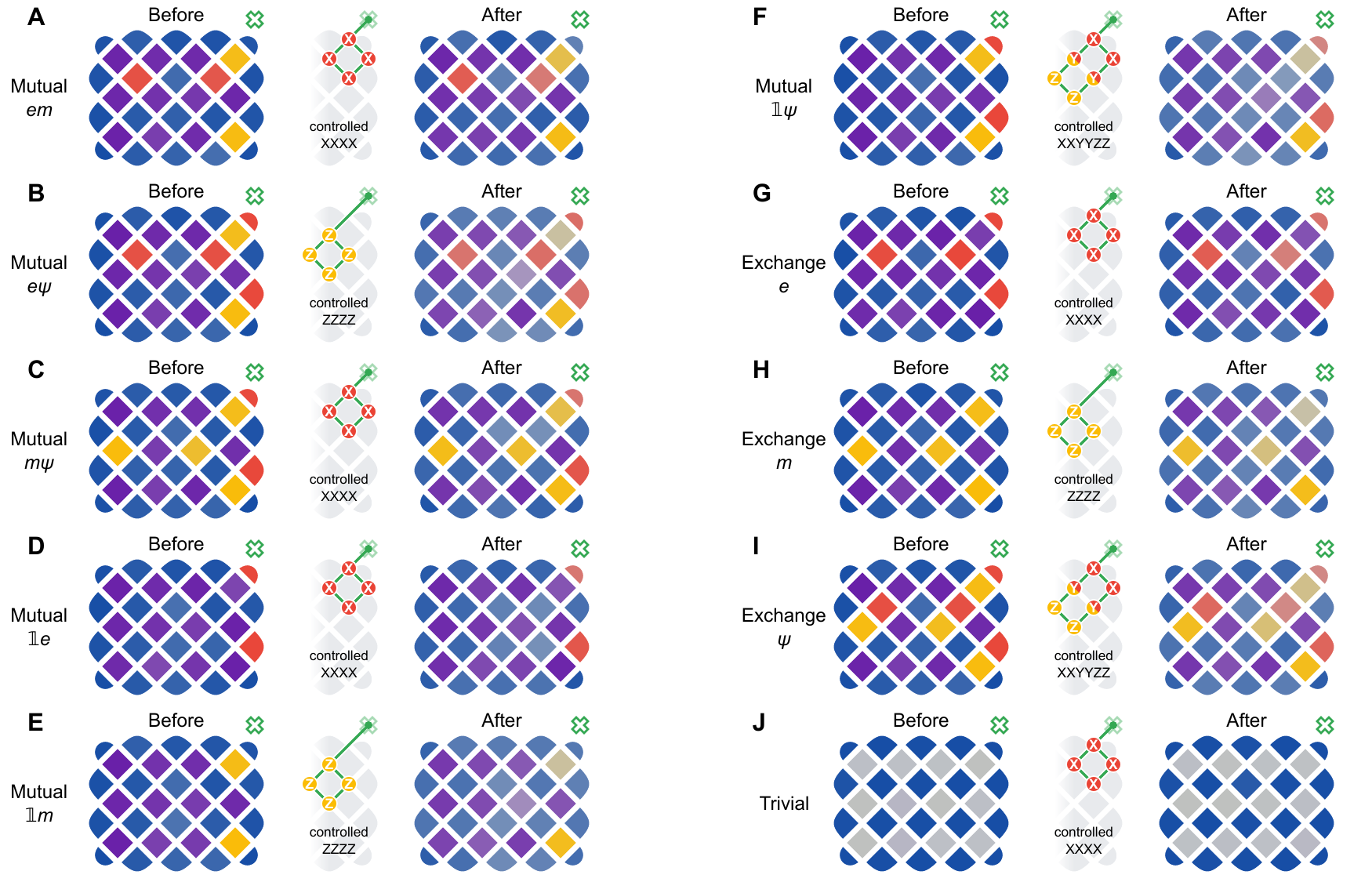}
   \caption{
   \textbf{Extended version of Fig.~3D-E.} 
   (\textbf{A-I}) Measured parity values for toric code eigenstates before and after the indicated controlled operation (green auxiliary qubit starts in $\ket{{+}}$), in the same order as the measured phases in Fig.~3F.
   (\textbf{A}) Same as Fig.~3D.
   (\textbf{B}) Note the simplification where two sets of \textit{XXXX} cancel, similar to the simplification in Fig.~\ref{fig:sm_fig3_decompositions}F.
   (\textbf{C}) Similar \textit{ZZZZ} cancellation.
   (\textbf{I}) Same as Fig.~3E.
   (\textbf{J}) Trivial case, effectively measuring $\bra{0000}XXXX\ket{0000}=0$ (also see Fig.~\ref{fig:sm_shotgun}C).
   }
   \label{fig:sm_fig3de_extended}
\end{figure*}

In our case, the unitary $U$ is a Pauli string simulating moving the anyons of the toric code. 
In Fig.~\ref{fig:sm_fig3de_extended}, we show the set of minimal experiments to measure all the braiding statistics between the anyons. Most of these paths can be understood based on the $\psi$ (fermion) exchange as shown in Fig.~\ref{fig:sm_fig3_decompositions}E. To exchange $\psi$, we first create two pairs of $\psi$ near the corner of the device.
Each movement of $\psi$ consists of a single Pauli $X$ and Pauli $Z$ that move the constituent $e$ and $m$ respectively. The resulting total path simplifies to a Pauli string \textit{XXYYZZ} (see Figure.~\ref{fig:sm_fig3_decompositions}F). Other minimal braiding paths can be deduced from the $\psi$-exchange case by only keeping the anyon of interest. 
The exception is the $e-m$ mutual statistics.
If we perform two exchanges between the $m$ near the corner and $e$ away from the corner, we can extract the $e-m$ mutual statistics with a Pauli string of 6 Pauli operators. However, a simpler path is to move $e$ around $m$ (topologically equivalent to exchanging $e$ and $m$ twice).   This path only consists of a Pauli string \textit{XXXX} (4 operators) as shown in Figure.~\ref{fig:sm_fig3de_extended}A.

The major cost of the procedure comes from the implementation of the controlled-$U$ that controls the auxiliary qubit and targets the support of $U$, which in general will involve multiple swap gates when decomposed into nearest-neighbor CNOTs. In order to reduce the depth of the circuit, we made use of a second auxiliary qubit (gray). The two auxiliary qubits are initially entangled in a Bell pair $(\ket{00}+\ket{11})\sqrt{2}$. Then we can parallelize the decomposition of the controlled-$U$ by using both auxiliary qubits as control qubits (see Figure.~\ref{fig:sm_fig3_decompositions}A-D. At the end of the circuit, we disentangle the second auxiliary qubit from the system by a single CNOT. This trick roughly halves the depth of the circuit in terms of nearest-neighbour gates. 
Rather than disentangling, it is also possible to directly measure a Pauli string $\ev{U}{\varphi} = \ev{X_aX_b}{\Psi} + i\ev{X_aY_b}{\Psi}$, where $a,b$ are the two auxiliary qubits. In our case, disentangling is advantageous because our CZ error ($\approx 0.005$) is much lower than our measurement error ($\approx 0.04$); see Sec.~\ref{sm:benchmarks}.

This interferometric protocol can be generalized beyond the Abelian braiding statistics of the toric code to measure the braiding statistics of other models, including some with non-Abelian braiding statistics supporting universal quantum computation~\cite{Liu2021}.  

\subsection{Randomized compiling}\label{sm:subsection:randomized_compiling}

\begin{figure*}
   \centering
   \includegraphics[width=504pt]{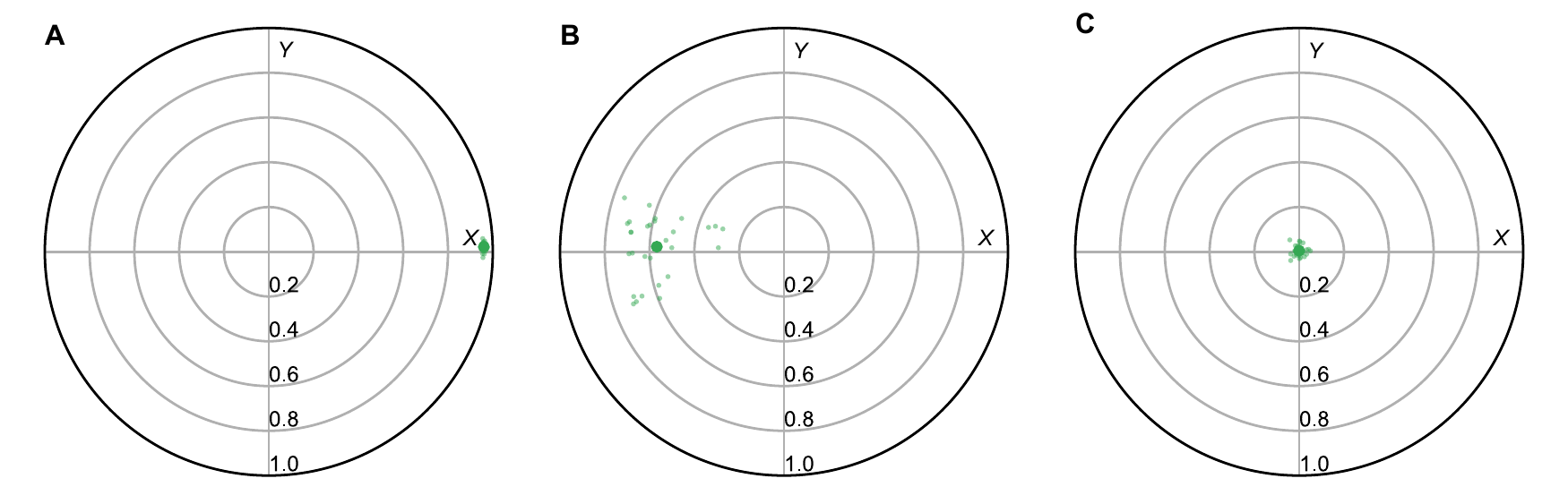}
   \caption{
   \textbf{Example scatter over randomized compiling instances.} 
   For each randomized compiling instance, we perform single-qubit tomography on the auxiliary qubit to obtain its Bloch vector.
   This single-qubit tomography consists of six sequences, effectively measuring along $\pm X$, $\pm Y$, and $\pm Z$, which averages out readout bias. We do not use any readout correction such as unfolding here.
   We plot the projection of the Bloch vector in the $XY$ plane, which determines the phase of the qubit state.
   The Bloch vector for each instance is shown in a smaller light green point (30 total), while the average Bloch vector over all the instances is a larger dark green point.
   (\textbf{A}) Control experiment where we prepare $\ket{+}$ and immediately perform tomography. The measured phase is $(0.007\pm 0.001)\pi$ (see text for discussion of estimating the phase uncertainty). The mean Bloch vector length is 0.96, where the discrepancy from 1.0 is dominated by measurement error.
   (\textbf{B}) Data used for the $em$ mutual datapoint in Fig.~3F, also connected to Fig.~3D and Fig.~\ref{fig:sm_fig3de_extended}A.
   Note the scatter in the data from individual instances, which we attribute to coherent and non-Markovian errors manifesting differently in different compiled instances.
   (\textbf{C}) Control experiment measuring \textit{XXXX} on the trivial state $\ket{0}^{\otimes 31}$, as shown in Fig.~\ref{fig:sm_fig3de_extended}J.
   We measure a Bloch vector close to (0, 0, 0) without any well-defined phase, as expected, since $\ket{0}^{\otimes 31}$ is not an eigenstate of \textit{XXXX}: $\bra{0000}XXXX\ket{0000}=0$.
   }
   \label{fig:sm_shotgun}
\end{figure*}

For the phase measurements in Fig.~3F, we utilize randomized compiling~\cite{Wallman2016, Beale2020}. This is a more sophisticated technique than the circuit optimizations described above (primarily inserting many $X$ gates) used for all the other experiments. Essentially, the layers of single-qubit gates (between layers of CZs) are transformed by random single-qubit Paulis in such a way that the overall circuit unitary is unchanged. We use 30 different randomly-compiled instances for each experiment. Each individual instance has a different perspective on the various coherent and non-Markovian errors that we wish to mitigate.

For example, in Fig.~\ref{fig:sm_shotgun}, we plot the scatter in the measured Bloch vector for the $em$ mutual measurement from Fig.~3F, as well as two control experiments.
The key result of Ref.~\cite{Wallman2016} is that by averaging over randomly-compiled instances, we tailor these coherent and non-Markovian errors into a depolarizing channel, which is suitable since here our focus is on extracting the \emph{phase} of a qubit after a sophisticated and deep 33-qubit circuit. The price is that all the errors now manifest incoherently, so the Bloch vector length is decreased.

There is not a well-established method of estimating the uncertainty in the phase, so we employ a simple technique, jackknife resampling~\cite{Tukey1958}. Resampling techniques are appealing here because each individual instance is subject to significant coherent and non-Markovian error, while averaging over many instances should be less sensitive. For each instance $i$ (of $n=30$ total), we compute the phase $\bar{\theta}_i$ averaging the Bloch vector over the $n-1$ \emph{other} instances. The average over all $n$ instances is $\bar{\theta}$. Then we estimate the standard error of the mean value of $\bar{\theta}$,
\begin{equation*}
    \sigma = \left[\frac{n-1}{n}\sum_{i=1}^n\left(\bar{\theta}_i-\bar{\theta}\right)^2\right]^{1/2}.
\end{equation*}
This is how we compute the error bars in Fig.~3F.

%
%
\section{Logical qubit states}\label{sm:logical}
\subsection{Logical state measurement}

Now we expand on the logical state measurement and error correction illustrated in Fig.~4B. 
The logical measurement proceeds as follows.  
We fix a basis, $Z$ or $X$, and measure all qubits in that basis.
We use the resulting bitstring to evaluate the logical operator, $Z_L$ or $X_L$, respectively.
At this stage, we work with individual measured bitstrings, rather than probability distributions. 
The bitstring can be used to evaluate the local parities, $A_s$ or $B_p$, respectively, equal to $\pm 1$.  
We perform error correction on the logical measurement by finding the minimal set of qubits to flip such that all local parities are $+1$.

There are various ways to choose which qubit measurements to flip.
Here, we use a brute force approach. 
Consider $Z$ basis.
For the distance-3 surface code, the logical $Z$ states $\ket{0_L}$ and $\ket{1_L}=X_L\ket{0_L}$ are superpositions of $2^4=16$ bitstrings.
For the distance-5 surface code, the logical $Z$ states are superpositions of $2^{12}=4096$ bitstrings.
Each bitstring satisfies $A_s=+1$ for all $s$.
By taking the Hamming distance (number of differing bits) between the measured bitstring and all the constituent bitstrings of the logical $Z$ states, we can find which constituent bitstring is closest to the measured bitstring.
The correct logical measurement outcome is then simply whether that closest bitstring is associated with $\ket{0_L}$ or $\ket{1_L}$.  
The logical $X$ measurement proceeds similarly with all measurements in the $X$ basis.

\begin{figure*}
   \centering
   \includegraphics[width=252pt]{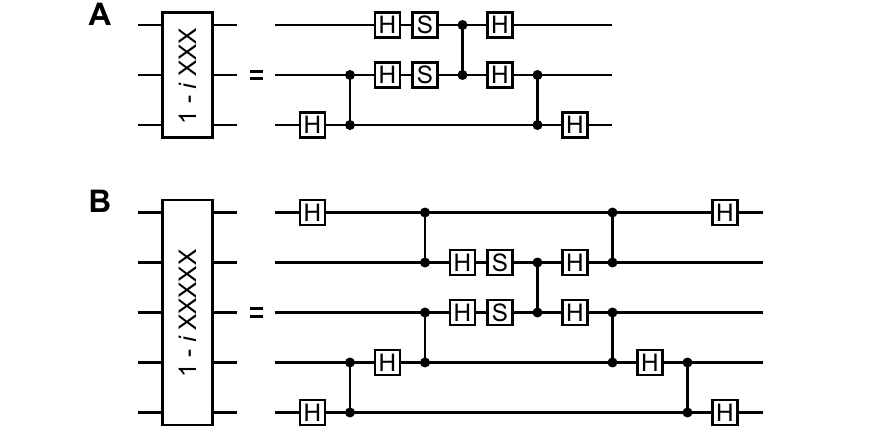}
   \caption{
   \textbf{CZ decomposition for $X_L^{1/2}$.} 
   The operator $X_L^{1/2}=(\mathbb{I}-iX_L)/\sqrt{2}$ is useful for logical tomography as it maps $\ket{+i_L}=(\ket{0_L}+i\ket{1_L})/\sqrt{2}\to\ket{0_L}$.
   We operate on the qubits that appear in $X_L$.
   The circuit decomposes the operator $(\mathbb{I}-iX_L)/\sqrt{2}$ into CZ, H, and S ($Z^{1/2}$) for the (\textbf{A}) $3\times 3$ and (\textbf{B}) $5\times 5$ qubit arrays.
   The S and H gates are compiled into one step, and we use the optimizations discussed in Fig.~\ref{sm:fig:sm_circuit_optimization}.
   }
   \label{fig:sm_xl_circuits}
\end{figure*}

As discussed in the main text, we use a logical operation $X_L^{1/2}$ for logical tomography. This rotates the $Y_L$ axis onto $Z_L$, so that measuring $Y_L$ on $\ket{\psi_L}$ is nominally equivalent to measuring $Z_L$ on $X_L^{1/2}\ket{\psi_L}$. Unfortunately, this is a nontrivial entangling operation that essentially involves shrinking the $X_L$ observable to two qubits, performing the desired rotation, and expanding the $X_L$ observable back across the array. This makes it especially vulnerable to errors, similar to the state injection protocol. This can be generalized for other powers $X_L^\alpha$ (and also $Z_L^\alpha$) using a complementary circuit.  The ladder structure also generalizes to larger code distances. We show the specific CZ circuits used for $X_L^{1/2}$ in Fig.~\ref{fig:sm_xl_circuits}.

\subsection{Dynamical decoupling}

Studying the onset of logical errors over time in Fig.~4D, we observe a significant basis dependence where $\ket{{+}_L}$ decays much more rapidly than $\ket{1_L}$. This is expected due to qubit frequency drift and low-frequency noise, which manifest as $Z$ errors. For example, if a qubit has a constant 500~kHz frequency offset, it will precess a $\pi$ rotation ($Z$ error) in 1~$\mu$s. These issues can be studied and mitigated using dynamical decoupling, techniques developed for nuclear magnetic resonance~\cite{Hahn1950, Carr1954, Meiboom1958, Gullion1990} that have been adopted successfully for superconducting qubits~\cite{Bylander2011}.

The dynamical decoupling we test in Fig.~4D is extremely simple. Given a particular wait time $t$, we apply an $X$ gate on each qubit at $t/4$ and again at $3t/4$, very similar to a ``spin echo" sequence. These $X$ gates nominally cancel out each other, but they also cancel out quasi-static $Z$ rotations over the course of the wait time. As shown in Figs.~\ref{fig:sm_logical_1us}-\ref{fig:sm_logical_5us}, using dynamical decoupling dramatically improves the performance for $\ket{{+}_L}$ and $\ket{{-}_L}$ ($X_L$ eigenstates sensitive to $Z$ errors), while it does not make a significant difference for $\ket{0_L}$ or $\ket{1_L}$, as expected.

Minimizing idle error in these states is important for the surface code, where an appreciable amount of time is spent idling while stabilizers are measured~\cite{Chen2021}. One direction to explore in future work is to look at other states, notably $Y_L$ eigenstates.
Developing protocols that work well for all logical states, such as alternating $X$ and $Y$ pulses (see Ref.~\cite{Gullion1990}) are highly desirable.
Another direction is to study the noise frequency spectrum by using different numbers of decoupling pulses, as demonstrated for a single physical qubit in Ref.~\cite{Bylander2011}.

\subsection{Extended experimental results}\label{sm:benchmarks}

\begin{figure*}
   \centering
   \includegraphics[width=504pt]{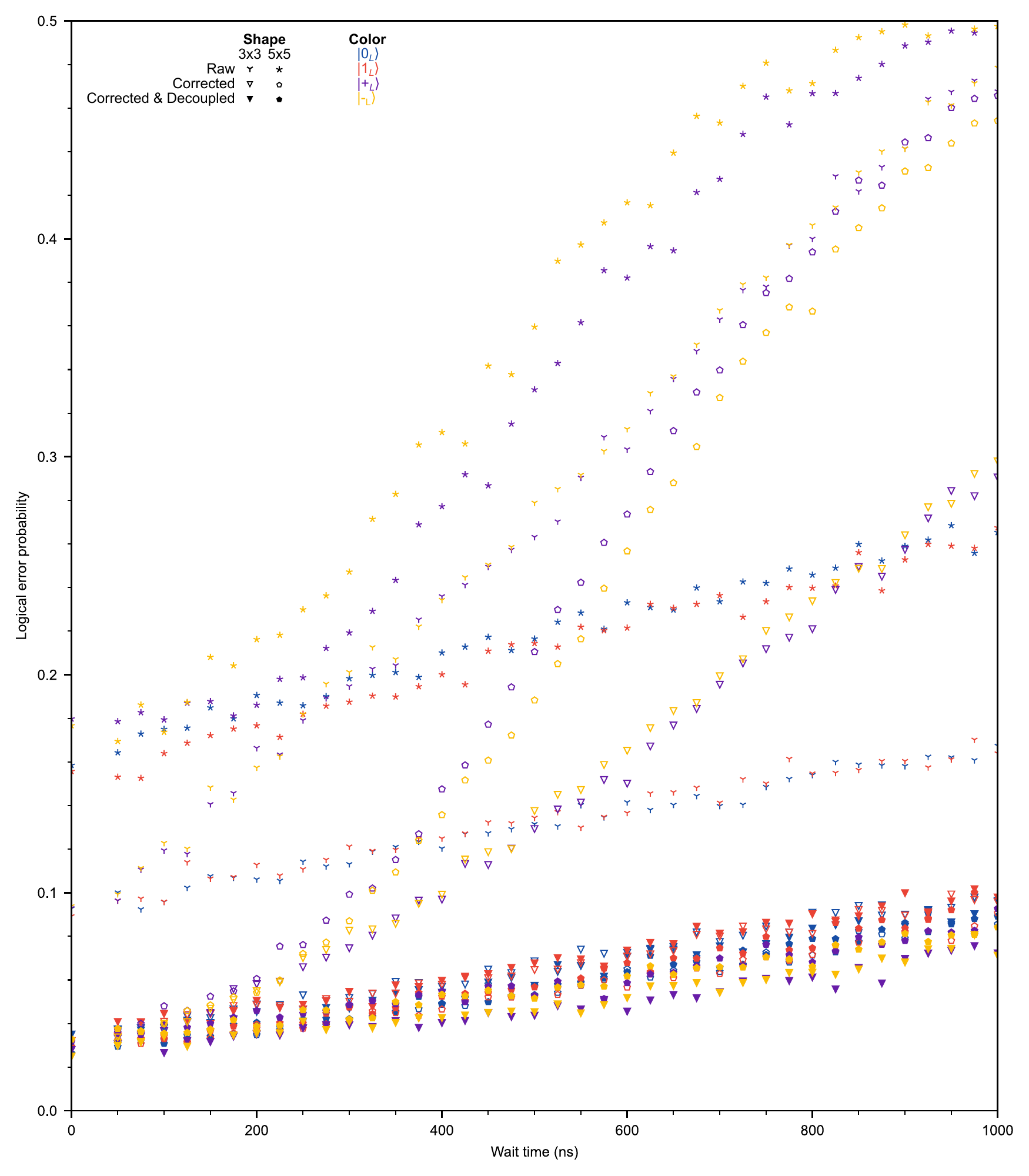}
   \caption{
   \textbf{Logical error versus wait time (1~$\mu$s).} 
   Extended version of Fig.~4D.
   }
   \label{fig:sm_logical_1us}
\end{figure*}

\begin{figure*}
   \centering
   \includegraphics[width=504pt]{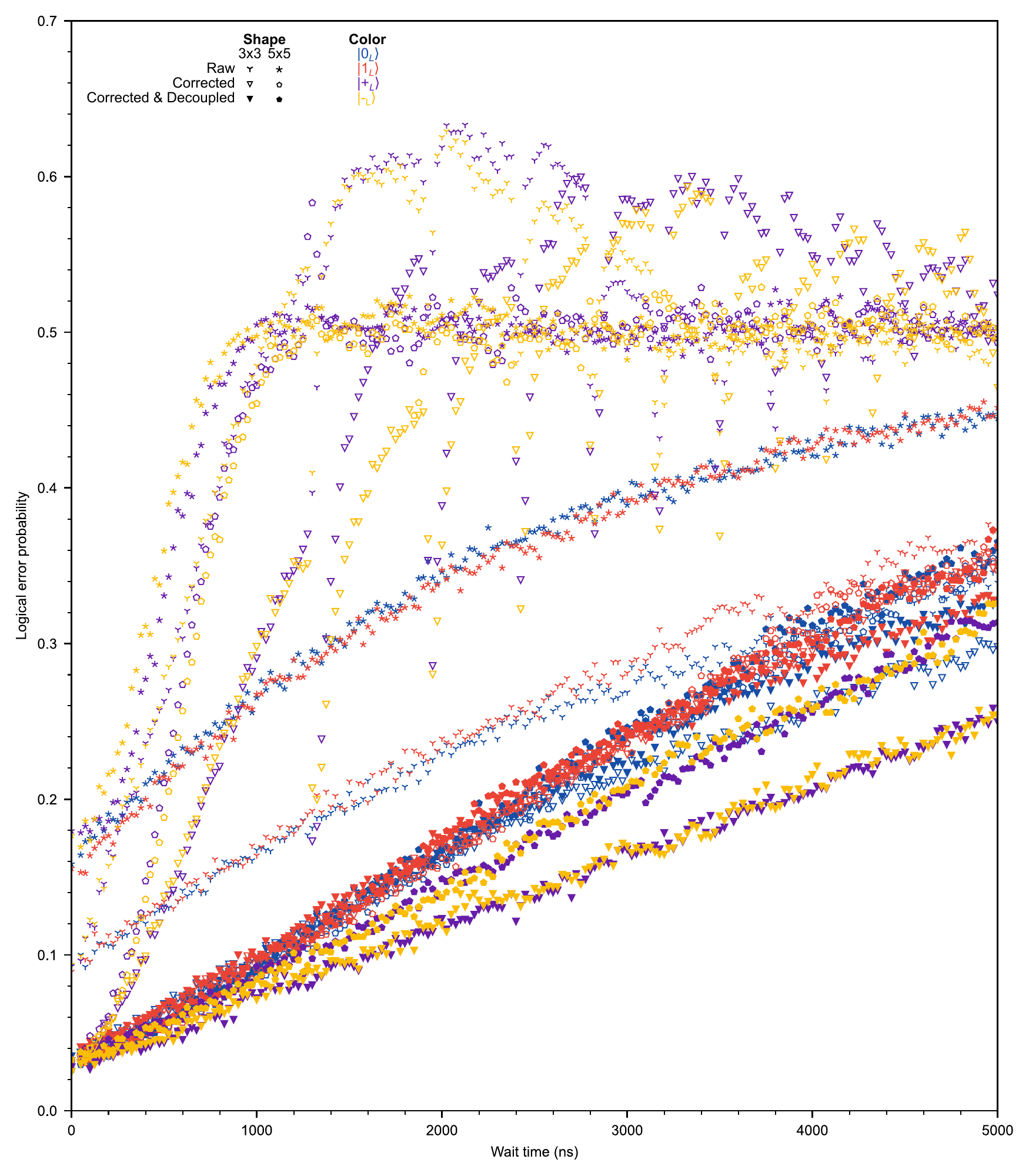}
   \caption{
   \textbf{Logical error versus wait time (5~$\mu$s).} 
   Extended version of Fig.~4D.
   }
   \label{fig:sm_logical_5us}
\end{figure*}

\begin{figure*}
   \centering
   \includegraphics[width=504pt]{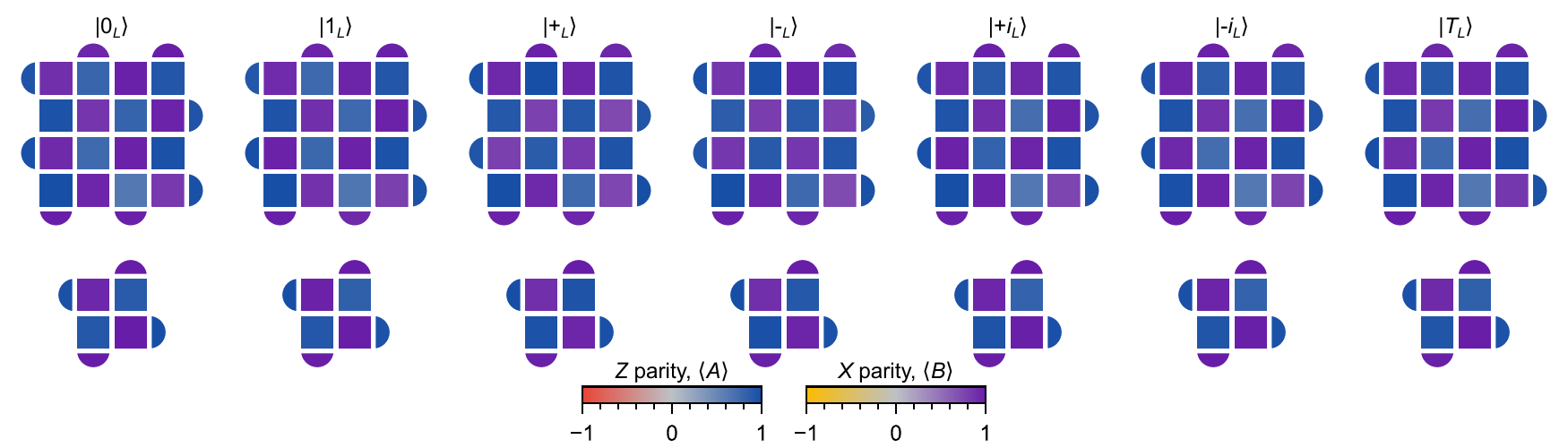}
   \caption{
   \textbf{Local parity measurements for various logical states.} 
   Extended version of Fig.~4A. Experimental parity measurements $\langle A_s\rangle$ and $\langle B_p\rangle$. For each column, we prepare a different logical state. We prepare $Z_L$ eigenstates ($\ket{0_L}$ and $\ket{1_L}$) and $X_L$ eigenstates ($\ket{{+}_L}$ and $\ket{{-}_L}$) directly. We prepare $Y_L$ eigenstates ($\ket{{+}i_L}$ and $\ket{{-}i_L}$) and $\ket{T_L}=(\ket{0} + e^{i\pi/4}\ket{1})/\sqrt{2}$ using state injection. Top row: $5\times 5$, bottom row: $3\times 3$. The rightmost column is the same data as Fig.~4A.
   }
   \label{fig:sm_logical_glass}
\end{figure*}

\begin{figure*}
   \centering
   \includegraphics[width=252pt]{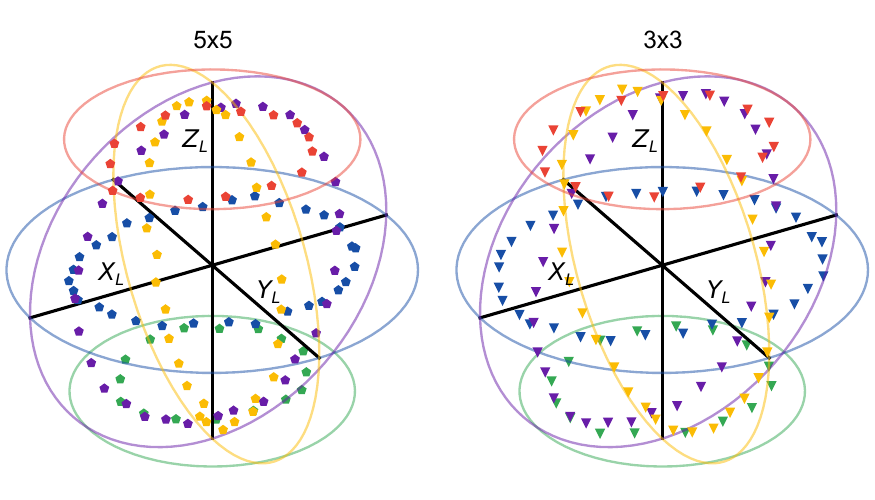}
   \caption{
   \textbf{Logical tomography of injected states.} 
   Extended version of Fig.~4C. Left: Logical tomography of injected states for the $5\times 5$ logical qubit, same as Fig.~4C. Right: $3\times 3$ version.
   }
   \label{fig:sm_bloch}
\end{figure*}

In Figs.~\ref{fig:sm_logical_1us}-\ref{fig:sm_logical_5us}, we present the data from Fig.~4D with extended context. We examine all four $Z_L$ and $X_L$ eigenstates for both $5\times 5$ and $3\times 3$, for raw measurement, corrected measurement, and corrected measurement with dynamical decoupling. Primarily, this supports the claim in the main text that dynamical decoupling does not substantially affect the $Z_L$ eigenstates $\ket{0_L}$ and $\ket{1_L}$. We observe that generally $\ket{0_L}$ and $\ket{1_L}$ ($Z_L$ eigenstates) behave similarly, as do $\ket{{+}_L}$ and $\ket{{-}_L}$ ($X_L$ eigenstates). Note the interesting oscillations for $\ket{{+}_L}$ and $\ket{{-}_L}$ without dynamical decoupling with microsecond timescale, only visible in Fig.~\ref{fig:sm_logical_5us}. The oscillations are most pronounced for the $3\times 3$ data and may come from individual qubits' static frequency offsets. The sharp dips in the corrected data suggest brief windows when the $Z$ errors coherently cancel enough that error correction can still succeed.

In Fig.~4A, we display local parity measurements for a particular logical state to illustrate we are in a toric code ground state (all local parities close to +1). In Fig.~\ref{fig:sm_logical_glass}, we plot similar data for seven different logical states, both for $5\times 5$ and $3\times 3$. Although the logical states can be distinguished by \emph{global} observables ($Z_L$ and $X_L$), they all look the same to the local parity operators $A_s$ and $B_p$. We also show logical tomography data for both $5\times 5$ (same as Fig.~4C) and $3\times 3$ state injection in Fig.~\ref{fig:sm_bloch}. Note the longer Bloch vectors for the $3\times 3$ case: the state injection, state preparation, and $Y_L$ tomography circuits are all lower depth for the $3\times 3$ case (each one has linear depth in code distance).

%
%
\section{Experimental details}

We use precisely the same Sycamore processor and experimental setup as in Ref.~\cite{Arute2019a}. We use CZ gates, resonantly swapping $\ket{11}$ with $\ket{02}$ and back; see Refs.~\cite{Foxen2020, Chen2021}. We optimize a frequency configuration for CZ gates with 35 active qubits and the others biased to low frequency, similar to Ref.~\cite{Chen2021}.

\begin{figure*}
   \centering
   \includegraphics[width=504pt]{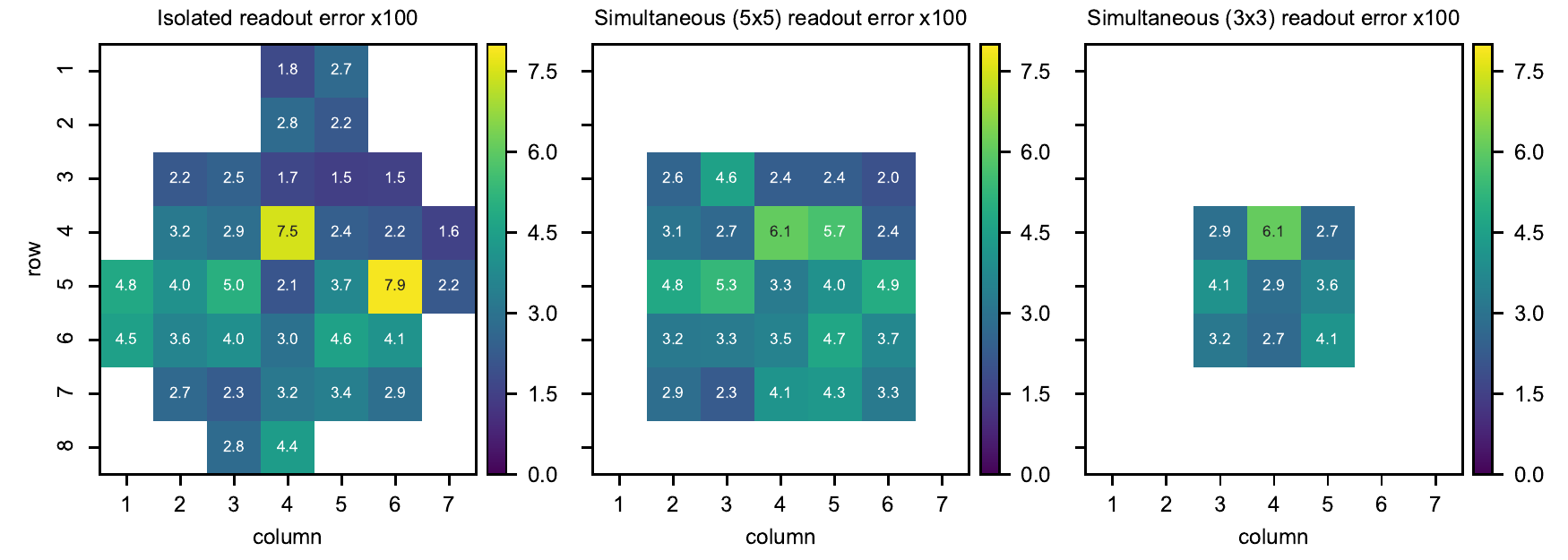}
   \caption{
   \textbf{Typical readout error.} 
   State preparation and measurement error on each qubit, averaging over $\ket{0}$ and $\ket{1}$ error.
   Left: ``Isolated" error measuring one qubit at a time.
   Center: Simultaneous 25-qubit error for the $5\times 5$ system used in Fig.~4.
   Right: Simultaneous 9-qubit error for the $3\times 3$ system used in Fig.~4.
   }
   \label{fig:sm_readout}
\end{figure*}

\begin{figure*}
   \centering
   \includegraphics[width=504pt]{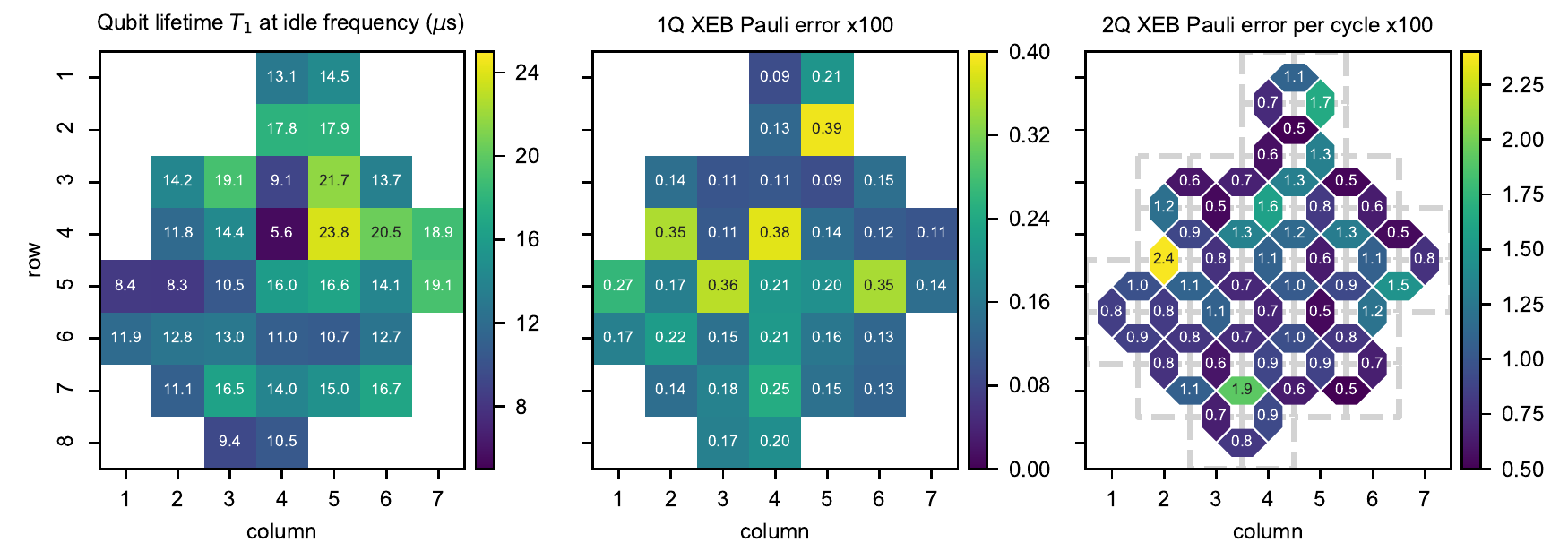}
   \caption{
   \textbf{Typical lifetime and gate error.} 
   Left: Qubit lifetime $T_1$ for each qubit, measured at its idle frequency in our configuration.
   Center: Single-qubit $\pi/2$ pulse cross-entropy benchmarking results, in Pauli error. Median: 0.0016.
   Right: Two-qubit CZ cross-entropy benchmarking results, in Pauli error per cycle. Each cycle contains one CZ and two single-qubit $\pi/2$ pulses. Median: 0.0084.
   }
   \label{fig:sm_xeb}
\end{figure*}

In Fig.~\ref{fig:sm_readout}-\ref{fig:sm_xeb}, we map experimental benchmarks across our qubit configuration. The center qubit, both in Fig.~1A and Fig.~4A, is (row, column) = (5, 4). The toric code rectangle (Fig.~1A) is rotated $45^\circ$ with respect to these plots. The auxiliary qubits used in Fig.~3 are (1, 4) and (1, 5). Qubits (3, 2) and (7, 6) are only used in the $5\times 5$ logical qubit experiments, Fig.~4.

In Fig.~\ref{fig:sm_readout}, we present typical readout error benchmark results. Each experiment involves readout assignment error and also state preparation error. Readout assignment error is dominant, for example from unwanted transmon transitions and separation error. State preparation error includes  stray $\ket{1}$ population, typically $< 0.01$, and $\pi$ pulse error, typically $\approx 0.001$.
The single-qubit measurements (left panel) are representative of the errors we experience in multi-qubit experiments. In the center and right panels, we show the specific errors we observed for simultaneous 25-qubit and 9-qubit readout used for logical measurements in Fig.~4. We benchmark 200 random bitstrings and 2000 repetitions each, then plot the fraction of runs where each qubit had an error. For more details about the readout setup, calibration, and benchmarking, see Ref.~\cite{Arute2019a}.

In Fig.~\ref{fig:sm_xeb}, we present typical qubit lifetime and gate error. We benchmark gate error (single-qubit $\pi/2$ pulses and two-qubit CZ gates) using cross-entropy benchmarking (XEB). Note we present Pauli error, and the CZ benchmarks (right panel) are error per cycle, where a cycle includes a CZ and two single-qubit $\pi/2$ pulses.
Taking into account the single-qubit gate errors, the typical CZ Pauli error is about 0.005.
In these benchmarks, we only examine one qubit or pair at a time, while we use many different gate patterns throughout the experiments in the main text. For more details, see Ref.~\cite{Arute2019a}.

\end{document}